\documentclass[aps,prb]{revtex4}
\usepackage{graphicx}
\tighten
\renewcommand{\vec}[1]{{\mathbf{#1}}}
\newcommand{\beq}{\begin{eqnarray}}
\newcommand{\eeq}{\end{eqnarray}}
\begin{document}
\draft

\title
{The Full Mottness}
\author{Tudor D. Stanescu}
\affiliation{Department of Physics and Astronomy, Rutgers University,
136 Frelinghuysen Rd., Piscataway, NJ., 08854-8019}
\author{ Philip Phillips}
\affiliation{Loomis Laboratory of Physics,
University of Illinois at Urbana-Champaign,
1100 W.Green St., Urbana, IL., 61801-3080}

\begin{abstract}
Though most fermionic Mott insulators order at low temperatures, 
ordering is ancillary to their insulating behaviour.
Our emphasis here is on disentangling ordering from the intrinsic strongly
correlated physics of a doped half-filled band.  To this
end, we focus on the 2D Hubbard model. 
Because the charge gap arises
from on-site correlations, we have been refining the non-perturbative
approach of Mancini and Matusumoto\cite{mm} which incorporates local physics. 
Crucial to this method is a self-consistent two-site dynamical
cluster expansion which builds in the nearest-neighbour energy scale, $J$.
  At half-filling,
we find that the spectral function possesses a gap of order $U$ and is devoid
 of 
any coherent quasi-particle peaks although ordering or charge fractionalization
are absent. At
 low temperatures,
local antiferromagnetic correlations emerge.  
In the doped case, we find that the Fermi surface exceeds the Luttinger volume.
The breakdown
of Luttinger's theorem in the underdoped regime
is traced both to the dynamically generated Mott gap as well
to a non-vanishing of the imaginary part of the self
energy at the Fermi level. Spectral weight transfer
across the Mott gap also emerges as a ubiquitous feature of 
a doped Mott insulator and suggests that high and low energy
scales are inseparable.   Additionally in the 
underdoped regime, we find that a pseudogap exists in the single
particle density of states as well as in the heat capacity.
 The pseudogap (which is set by the
energy scale $t^2/U$) is argued to be
 a ubiquitous feature of a lightly
doped Mott state and simply represents the fact that hole transport
involves double occupancy.  In analogy with
 the Mott gap and antiferromagnetism,
we propose that ordering may also accompany the formation
of a pseudogap.  We suggest a current pattern
within a 1-band model that preserves translational but
breaks time-reversal symmetry along the canonical x and y axes but not along
$x=\pm y$ that is consistent with the experimental observations.  
Finally, we show that the Hall coefficient in 
a doped Mott insulator must change sign
at a doping level $x<1/3$.  The sign change is tied to a termination of strong correlation physics
in the doped Mott state.
\end{abstract}

\maketitle

\section{Introduction}

Electronic systems with an odd number of particles per unit cell
are typically metallic at zero temperature.  However, Mott\cite{mott} proposed
that such systems in the presence of strong interactions can insulate 
at zero temperature without any accompanying
symmetry breaking process, such as antiferromagnetism or charge ordering, which
necessarily double the unit cell.  Strictly then, a Mott insulator
(should it exist) is
a paramagnetic state with an odd number of particles per unit cell. 
Insulating behavior arises from the charge gap generated by
the projective mismatch
between the
sub-lattices
which have zero or some finite fraction of doubly occupied sites.  
When the overlap between such sub-lattices is sufficiently small, 
no transport obtains. 
In bosonic systems with a single boson per site, a true zero temperature 
Mott insulating
state is realized in the standard quantum rotor model or 
Bose-Hubbard model when the on-site charging energy
exceeds a critical value such that phase coherence is destroyed\cite{bh1,bh2,bh3}.  In fact, the recent
observation\cite{bloch} that a Bose condensate in an optical lattice 
can be tuned between a superfluid and a Mott insulator
simply by changing the intensity of the laser light places the bosonic Mott
insulator on firm experimental footing.  

However, for fermionic systems, the inherent problem with the Mott insulating state of a half-filled band in $d>1$ is its proclivity
to order at zero temperature.   Consequently, it is tempting to 
equate the Mott insulator with the ordered state or to assert
categorically that the Mott insulator as an entity distinct from
a symmetry-broken state never 
existed in the first place.  In the context of high $T_c$ in the cuprates,
which are all anti-ferromagnetic Mott insulators, both 
views\cite{laughlin} have been 
strongly expressed.   In fact, a large body of work
on the cuprates has focused primarily on models
that capture low-energy spin physics\cite{s1,s2,s3} at or near an antiferromagnet or a charge-ordered state\cite{c1,c2,c3,c33,c4,c5,c6,c7,c8}, or a 
classification\cite{dhlee,sachdevclas} of the various charge-ordered states that ensue in a half-filled band. 
Alternatively, low-energy spin liquid models\cite{anderson,sl1,sl2,sl3,sl4,sl5,sl6,sl7} 
(that is, models
with spin translation and spin rotation symmetry) have been proposed as 
candidates to describe the Mott insulator. In such approaches,
the high energy scale associated with the charge gap is argued
to be irrelevant, hence the focus exclusively
on the spin sector to characterise the Mott insulator.  Of  course such an approach presupposes
that the high and low energy degrees of freedom can be
disentangled.  

Should the insulating state in a half-filled band prove to be
nothing other than a mean-field broken-symmetry state, then
fermionic Mott insulators do not exist.  
Hence, a relevant question for the Mott insulating state is as follows:  If we subtract the fact that
ordering obtains at zero temperature, is there anything left over
that is not explained by ordering?  Equivalently, does ordering 
provide an exhaustive explanation of the state proposed by Mott?  We refer to
whatever might be left over once we subtract the fact that ordering has occurred
as {\it Mottness}.  
That something might be left over is immediately
evident from the fact that charge and spin ordered states all result in a doubling of the unit cell
and hence are adiabatically connected
to an insulator with an even number of electrons per unit cell. A typical example of such a system is a band insulator.  Contrastly, Mott insulators which have an odd number of
electrons per unit cell are not.  
Additionally, spin and charge ordered states 
can be described at the level of Hartree-Fock simply by
constructing the correct broken symmetry ansatz.  However the Mott state
has no weak coupling or Hartree-Fock counterpart.  Hence, although the Mott state might
be unstable to ordering at low temperature, some features should be left over
which represent the fingerprint of the 
nonadiabaticity with a band insulator 
and the fact that this state arises fundamentally from
 strong electron correlation.
For example, experimentally\cite{cooper,uchida1}, it is clear that above any
temperature associated with
ordering in both the electron and hole-doped cuprates, a charge gap of order 2$eV$ is present in the optical conductivity and oxygen K-edge photoemission\cite{chenbatlogg}.   Hence, the 
vanishing of the low-energy ($<1eV$) spectral weight at high temperature is not linked
to magnetism or ordering of any kind.  Further, the
electronic bands below and above the charge gap are not rigid as would be the 
case in a band insulator.   To illustrate, as a function of doping, in both hole and electron-doped
cuprates\cite{cooper,uchida1}, the 
low-energy spectral weight increases at the expense of the high-energy ($>2eV$)
spectral weight such that the total integrated
optical conductivity remains constant up to $4eV$.  The same massive reshuffling of spectral weight from 2eV above the Fermi energy is also observed in 1-particle probes such as oxygen K-edge photoemission\cite{chenbatlogg} and angle-resolved photo-electron
spectroscopy (ARPES)\cite{ncco,cuclo}.  Such
spectral weight transfer indicates that the total number of low-energy 
degrees of freedom in the normal state of the cuprates cannot be decoupled from the high energy scales.  What is surprising about the cuprates is that even when superconductivity obtains, the low and high energy degrees of freedom are still coupled.
For example,  R\"ubhaussen, et. al.\cite{rubhaussen} have shown
that changes in the optical conductivity occur at energies 3eV (roughly 
100$\Delta$, $\Delta$ the maximum superconducting gap) away from the 
Fermi energy at $T_c$, and Bolegr\"af, et. al.\cite{marel} have seen an 
acceleration in the depletion of the high energy spectral weight accompanied 
with a compensating increase in the low-energy spectral weight at and below the
superconducting transition.  Similarly, Bontemps, et. al.\cite{bontemps}  have directly 
observed that
in underdoped (but not overdoped) BSCO, the Glover-Ferrel-Tinkham sum rule is violated and the optical
conductivity must be integrated to $20,000cm^{-1}$ to recover the spectral
weight lost upon condensation into the superconducting state.  In a standard
BCS superconductor, condensation leads to loss of spectral weight at energy scales no more than ten times the pairing energy.  The fact that pair condensation
perturbs the optical conductivity on energy scales as large as 100$\Delta$ suggests
that there is a direct link between 
the high
energy Mott scale and superconductivity.  Further, the persistence of 
spectral weight transfer through the superconducting transition indicates
that the properties of 
the Mott state remain intact even in the presence of ordering.  While
spin-density wave antiferromagnets also possess two bands with a gap, such a
state is insufficient to explain the origin of the spectral weight transfer in
the cuprates.  The reason is simple: spectral weight transfer persists well 
above $T_N$ and at a doping level $\gg 2\%$, where
 antiferromagnetism is 
absent. 

Concretely, what the optical
conductivity and oxygen k-edge photoemission experiments on the cuprates
lay plain is that regardless of whether ordering obtains at sufficiently
low temperatures, the charge degrees of freedom remain characterized by a distinctly different state than that of any
ordered state or a band insulator.  To understand what physics is entailed
by this state, consider the Mott mechanism for generating an insulating
state in the one-band
Hubbard model.  At half filling, the chemical 
potential lies in the middle of 
the gap separating the lower and upper Hubbard bands which are dynamically
split by the on-site energy for double occupancy.   However, half the spectral weight
resides in each of the bands.  Consequently, to satisfy the sum rule that each state in the first Brillouin Zone (FBZ) carries
unit spectral weight,
the spectral function must be integrated across the charge gap not simply
up to the chemical potential.  Hence, 
the half-filled state is characterized by the Fermi energy lying in a gap but partially
occupied states exist.  
It is this seemingly contradictory state of affairs that is at the heart of 
Mottness.  Spectral weight transfer cannot obtain without it.  For example,
if each state below the chemical potential had unit spectral weight, no state would be available for spectral weight transfer from high energies.  
As a 
consequence, adding or removing an electron cannot be done without affecting
both high and low energy scales.  
Consequently, at any doping level,
the electronic states describing the charge carriers can be written
as linear combinations of excitations living in both the LHB and UHB as will be
detailed below. 
As a result, in the Mott state, the traditional notion that the chemical 
potential demarcates the boundary between
zero and unit occupancy fails fundamentally.  
Of course, in Fermi liquids, the spectral function for each $\vec k-$state
can also have an incoherent background which can extend to high energies.
However, as long as a coherence peak exists, a sharp criterion exists for unit occupancy
of each state, namely whether or not the coherence peak crosses the Fermi level. In a Mott insulator, no such coherence peak exists and consequently, incoherence dominates
the Mott state.

\begin{figure}
\centering
\includegraphics[height=7.5cm]{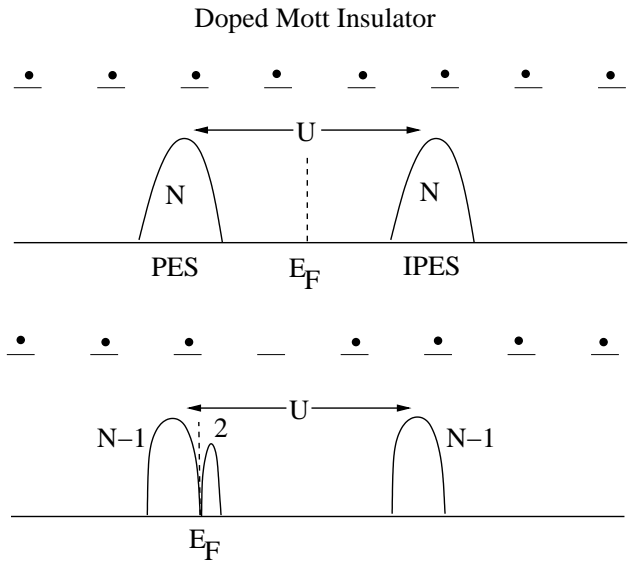}
\caption{Spectral weight transfer in a doped Mott insulator.  The Photoelectron
spectrum (PES) denotes the electron removal states while the electron-addition
states are located in the inverse photo-electron spectrum (IPES).  The on-site
charging energy is $U$. Removal of a single electron results in the creation
of two single particle states at the top of the lower Hubbard band. 
By state conservation, one state comes from the lower 
 and the other from the upper Hubbard band and hence spectral weight transfer across
the Mott gap. }
\label{fig1}
\end{figure}

Alternatively, one can view the spectral
weight transfer in real space by simply counting the number of 
available states for the photo-emission and inverse photo-emission spectra 
as demonstrated by Meinders, Eskes, and Sawatzky\cite{sawa}. We recount the argument here as it is simple and instructive.  Consider 
the half-filled one-dimensional chain of one-electron atoms shown in
Fig. (\ref{fig1}). 
Both the electron removal 
(photo-electron) and electron-addition (inverse photo-emission) spectral weight 
are equal to $N$ because at half-filling there are $N$ ways of adding or subtracting an electron from a site that is singly occupied.  When a single hole is added, both the electron-removal and the 
electron-addition spectral decrease to $N-1$ as there are now $N-1$ ways to 
add or subtract an electron
from sites that are already occupied.  Hence, there are two less states.  
These states correspond to the spin up and spin down states of the empty site 
and hence belong to the LHB.  
Consequently, the low-energy spectral weight (LESW), $\Lambda(x)$, has increased
 by two states.  One of the states must come from the UHB as
 the high energy part
 now has a spectral weight of $N-1$.  Hence, for a single hole,
there is a net transfer of one state from high to low energy.  
This argument is simply the real-space restatement of the more general
principle that in $\vec k-$space the LHB and UHB are not static but dynamic
and hence necessarily give rise to spectral weight transfer.  In general, 
simple state counting yields $2x$ for the growth of $\Lambda(x)$ and $1-x$
 for the depletion of the high energy sector.  In actuality, the dynamical contribution
to the LESW results in $\Lambda(x)>2x$.   The dynamical LESW corresponds to 
virtual excitations to the UHB.   Hence, in a strongly correlated system, the 
phase space available to add a single electron exceeds the nominal number of 
states initially present in the low energy scale, leading thereby to an inseparability of the low and high energy scales.
 
The ubiquity of spectral weight transfer in the cuprates 
places
extreme restrictions on which low-energy models are valid.  Consider the 
standard $t-J$ model\cite{anderson}.  As the $t-J$ model projects onto the LHB, we can estimate
the corresponding LESW exactly by counting the electron-removal states.
Consequently, $\Lambda(x)=2x$ is exact in the $t-J$ model.  
However, in the actual Hubbard model,
$\Lambda(x)>2x$.  Consequently, the standard $t-J$ model does not have the
correct number of low-energy degrees of freedom to describe the low-energy 
physics.  This problem can be corrected\cite{eskes}, however, by using the full
 strong-coupling
Hamiltonian that results from the $t^2/U$ expansion and replacing all the 
electron operators by their projected counterparts.  However, the price one
pays is that the new
projected electron operators do not obey the standard fermion commutation
relations.  That a correct low-energy theory must give up on either standard 
fermion commutation relations or particle conservation has already been pointed
 out by Meinders, Eskes, and Sawatzky\cite{sawa}.   Hence, all standard low-energy
fermionic theories do not have the correct physics to describe the cuprates
primarily because the low-energy degrees of freedom are not fermionic.   

Given the ubiquity of spectral weight transfer over a large part of
the phase diagram of the cuprates, it is imperative that any theory of high
$T_c$ incorporate the high energy scale associated with the 
charge gap. Further, because ordering does not seem to be a requirement
for spectral weight transfer, our focus is on an accurate description of the 
high-temperature charge vacuum of the doped Mott insulating state.  It is crucial
in such an approach that the hierarchy of energy scales, $U$, $t$, $t^2/U$, for example, emerge. 
Consequently, we have been refining the non-perturbative approach of Matsumoto and Mancini\cite{mm} to describe
the inter-relation between the energy scales in a doped Mott insulator.   This approach
is based on the experimental observation that the low energy scales in
doped Mott insulators are derived from high energies.  As a result, the beginning
point, namely the Hubbard operators, is one that is well-known to yield the Mott charge gap scale.  Successively smaller energy scales are derived by treating local correlations on a small cluster.  The self-energy of the lattice is then determined self-consistently from the local impurity problem.   The Matsumoto-Mancini\cite{mm} approach is then 
in the spirit of the cellular dynamical mean field treatment\cite{cellular}
We report here the full details of this 
approach and catalogue the general results that follow from Mottness.  Aside 
from a LESW that exceeds the nominal value obtained from state
 counting, we find that
1) Mottness gives rise to broad spectral features in the underdoped regime,
2) a violation of Luttinger's theorem in the underdoped regime,
3) hole-like Fermi surface near half-filling, 5) a jump in the chemical 
potential upon doping and 6) a pseudogap in the underdoped regime without 
invoking
any symmetry breaking. In general, we find that the pseudogap is due strictly to near-neighbour correlations and can be thought of as the nearest-neighbour analogue of the on-site generated Mott gap.  

\section{Method}

Many years ago, Hubbard\cite{hubbard} wrote the electron annihilation operator, $c_{i\sigma}$ as a linear combination
\beq
c_{i\sigma}=\eta_{i\sigma}+\xi_{i\sigma}
\eeq
of two composite excitations that reflect the energetically challenged
landscape an electron must 
traverse in the presence of a large on-site Coulomb repulsion, $U$. Physically, the operators $\eta_{i\sigma}=c_{i\sigma}n_{i-\sigma}$ and $\xi_{i\sigma}=c_{i\sigma}(1-n_{i-\sigma})$ represent an electron moving on doubly and singly
occupied sites, respectively.  Because such sites are split by $U$, the Hubbard operators lead naturally to a gap at half-filling in a paramagnetic state, a result which thus far,
only two other methods, dynamical mean-field theories (DMFT)\cite{dmft} and quantum Monte Carlo (QMC)\cite{qmc,scalapino}, have been able to obtain.  
Regardless of this success, the Hubbard operators have been criticized extensively because untested approximations generally accompany their implementation, and they lead to a Fermi surface which does not preserve the Luttinger volume.  However,
methods that are deemed as reliable\cite{puttika,ogata,jarrell2,qmc2}, also find that in the lightly doped regime, Luttinger's theorem is violated as is seen experimentally\cite{ncco}.  Hence, a violation of Luttinger's theorem is not an {\it a priori} reason to
dismiss the Hubbard operators. In fact, any violation of Luttinger's theorem must occur in the lightly doped or non perturbative regime.  The untested or uncontrolled approximations usually arise from truncations in the equations of motion.  However, such problems can be circumvented by the following procedure. First, project all new operators that arise from the Heisenberg equations of motion
of the Hubbard operators onto the Hubbard basis.  Second, write the 
self energy exactly in terms of the remaining operators which are now orthogonal
to the Hubbard basis.  Third, use local DMFT methods to
calculate the resultant electron self energy.  The approximation introduced in
the third step is that the self energy for a finite cluster is used to
determine the self-energy for the interacting lattice.  However, such
methods have been shown to be strongly convergent and in fact constitute
the accepted methodology for treating strongly correlated systems.  In principle, as the cluster size is extended to infinity, such a cluster procedure becomes
exact.  Hence, an implementation of the Hubbard operators coupled with DMFT-type technology places the limitations not on truncation in the equations of motion
but on the accuracy of the impurity solver and the size of the finite cluster. 
It is such a procedure that we outline here.  As many of the details have been
left out in previous presentation in the original paper by
Mancini and Matsumoto\cite{mm} on the two-site cluster and subsequent implementations\cite{stanescu1}, we will provide a complete derivation of the method so that any one reading this paper can implement
it immediately.  The key features of this method are its ability to describe physics
on the scale of the Mott gap, $U$, as well as on the the scale $J\approx t^2/U$.

\subsection{Dynamical Green Function}

Our starting point is the on-site Hubbard model
\begin{equation}
H = -\sum_{i,j,\sigma} t_{ij}c_{i\sigma}^{\dagger}c_{j\sigma} + 
U\sum_{i} n_{i\uparrow}n_{i\downarrow}
\end{equation}
with nearest-neighbor hopping, $t_{ij} = t\alpha_{ij}$. We also introduce the composite operator basis
\beq
\psi_\sigma(i)=\left(\begin{array}{l}
\xi_{i\sigma}\\\eta_{i\sigma}
\end{array}\right)
\eeq
and its associated retarded Green function $S(i,j,t,t')  =  
\langle\langle \psi_{i\sigma};\psi^\dagger_{j\sigma}\rangle\rangle  =  
\theta(t-t')\langle \{\psi_{i\sigma}(t),\psi^\dagger_{j\sigma}(t')\}\rangle$.
Writing the equations of motion for the Hubbard basis and
projecting with the Roth\cite{roth} projector ${\cal P}(O)=\sum_{ln}\langle\{O,\psi_l^\dagger\}\rangle I_{ln}^{-1}\psi_n$, we obtain for the `current' operator the expression
\begin{equation}
j_i(t)=i\frac{\partial}{\partial t} \psi_i = K\psi_i+{\cal P}(\delta j_i)+\delta J_i=E\psi_i+\delta J_i.
\end{equation}
The formal solution for the Green function in Fourier space,
\begin{equation}
S({\bf k},\omega)=\frac{1}{\omega-{\bf E}({\bf k})-\delta {\bf m}({\bf k},\omega){\bf I}^{-1}({\bf  k})}{\bf I}({\bf  k})  \label{greenEq},
\end{equation}
contains the self-energy $\Sigma({\bf k},\omega) = \delta {\bf m}({\bf k},\omega){\bf I}^{-1}$ with
\begin{equation}
\delta {\bf m}({\bf k},\omega)=FT\langle \theta(t-t') \{\delta J (t),\delta J^\dagger (t')\}\rangle_I
\end{equation}
where the subscript $I$ indicates the irreducible part and $FT$ denotes the Fourier transform. 
We consider the paramagnetic case, for which the overlap matrix ${\bf I}$ is diagonal with $I_{11}\equiv I_1 = 1-n/2$ and $ I_{22}\equiv I_2 = n/2$.

The primary operational hurdle is the evaluation of the dynamical correction 
$\delta {\bf m}$.  The first step is to write the explicit expressions for the operators $\delta J$ which are 'orthogonal' to the basis $\psi$. Using the notation $\tilde{t} = 2d\cdot t$, $d$ being the dimensionality of the system,  and $\tilde{p} = p-n^2/4$, we obtain
\begin{equation}
\delta J_1(i) = -\tilde{t}\left[\pi_{i\sigma} + \frac{n}{2}c_{i\sigma}^{\alpha} - e(\xi_{i\sigma}I_1^{-1} - \eta_{i\sigma}I_2^{-1}) 
- \tilde{p}(\xi_{i\sigma}^{\alpha}I_1^{-1} - \eta_{i\sigma}^{\alpha}I_2^{-1}) 
\right]  \label{delphi}
\end{equation}
and $\delta J_2(i) = -\delta J_1(i)$, where the self-consistent parameters
 $e$ and $p = \tilde{p}+n^2/4$ as well as the higher order composite operator $\pi_i$ are given by
\beq\label{e}
e &=& \langle\xi_i^{\alpha}\xi_i^{\dagger}\rangle - 
\langle\eta_i^{\alpha}\eta_i^{\dagger}\rangle,\nonumber\\
p& = &\langle n_{i\sigma}n_{i\sigma}^{\alpha}\rangle + 
\langle c_{i\uparrow}^{\dagger}c_{i\downarrow}
(c_{i\downarrow}^{\dagger}c_{i\uparrow})^{\alpha}\rangle - 
\langle c_{i\uparrow}c_{i\downarrow}(c_{i\downarrow}^{\dagger}
c^{\dagger}_{i\uparrow})^{\alpha}\rangle,\nonumber\\
\tilde{t}\left(\begin{array}{c}\pi_{i\uparrow} \\
 \pi_{i\downarrow}\end{array}\right)& = &
\sum_{j}t_{ij}\left(\begin{array}{c} -n_{i\downarrow}c_{j\uparrow} 
+ c_{i\downarrow}^{\dagger}c_{i\uparrow}c_{j\downarrow} -
 c_{i\uparrow}c_{i\downarrow}c_{j\downarrow}^{\dagger} 
\\ -n_{i\uparrow}c_{j\downarrow} + 
c_{i\uparrow}^{\dagger}c_{i\downarrow}c_{j\uparrow} +
 c_{i\uparrow}c_{i\downarrow}c_{j\uparrow}^{\dagger}\end{array}\right). 
\eeq 
 and the superscript $\alpha$ denotes the averaging over nearest-neighbor sites. Consequently, we can write the dynamical correction matrix $\delta {\bf m}$ in the form
\begin{equation}
\delta {\bf m}({\bf k},\omega) = Dm({\bf k},\omega)\left(\begin{array}{cc}
1& -1\\ 
-1& 1\\
\end{array}
\right)
\end{equation}
and the problem reduces to the determination of the higher-order Green function $Dm({\bf k}) = FT\langle \theta(t-t')\{\delta J_n(t)\delta J_m^\dagger(t)\}\rangle_I$.
Because $Dm({\bf k},\omega)$ cannot be evaluated exactly, we seek a systematic
way of calculating the dynamical corrections.  
The simplest approach
would be to consider the single-site approximation.  Such an approximation
is in the spirit to the $d=\infty$ \cite{dmft} methods, in which
the self-energy is momentum independent.  An improvement would be to consider
the dynamics associated with two sites as proposed by Mancini
 and Matsumoto\cite{mm}. 
Evaluation of the self-energy over successively larger clusters
would lead to an exact determination of the dynamical corrections.
The essence of this approach is based on  the fact that the  physics of
 strongly correlated  electrons emerges mainly from local correlations: on-site
interactions generate the Mott gap, while nearest-neighbour interactions
generate the scale, $t^2/U$.  e must treat all excitations on equal footingSuccessively larger clusters build in lower and lower energy scales.  However, due to the fact that transfer of spectral weight from high to low energies is characteristic of Mott insulators, a separation of  energy scales in strongly correlated problems is not possible.  Therefore, it is crucial that all the excitations be treated on equal footing and, consequently, local correlations must be included. 
Following Mancini and Matsumoto\cite{mm}, we write the dynamical corrections as a series
\begin{equation}
Dm(x,x')=\delta_{x,x'}Dm_0(x,x')+\sum_a\delta_{x+a,x'}Dm_1(x,x')
+\cdots
\end{equation}
in increasing cluster size. 
Here, $x$ and $x'$ are two representative sites and $a$ indexes
all nearest-neighbor sites.  In the two-site approximation, the series is 
truncated at the level
of on-site, $Dm_0$, and nearest-neighbor, $Dm_1$ contributions. 
In Fourier space, the dynamical corrections can be written
as,
\begin{equation}
Dm({\bf k}, \omega)\approx Dm_0(\omega)+\alpha({\bf k})Dm_1(\omega).
\end{equation}
Here $Dm_0$ and $Dm_1$ involve Green functions of operators defined on nearest neighbor sites (see Eq. \ref{delphi}). Consequently, these Green functions
 contain operators defined on up to four sites.
Further simplifications can be made if we assume that the dominant contributions arise from terms involving  at most two  nearest-neighbor sites, $x$ and $x'$. In this case, the superscript $\alpha$ in Eq. (\ref{delphi}) corresponds to a particular neighboring site instead of an average over all the nearest-neighbor sites. With this assumption, we obtain 
\begin{eqnarray}
Dm_0(\omega)&=&\frac{1}{2d}FT\langle \theta(t-t')\{\delta J(t),\delta J^\dagger (t')\}\rangle_I,\nonumber\\
Dm_1(\omega)&=&\frac{1}{2d}FT
\langle \theta(t-t')\{\delta J(t),\delta J^{\prime\dagger} (t')\}\rangle_I,                \label{DmEq}
\end{eqnarray}
where the factor of $1/2d$ arises from the coordination number,
$\alpha({\bf k})=\frac{1}{d}\sum_l~\cos(k_l)$, and $\delta J$ and 
$\delta J^{\prime}$ are centered
on two nearest-neighbor sites, $x$ and $x'$, respectively.
It is straightforward now to express the Green function in terms of these quantities,
\begin{equation}
S({\bf k}, \omega) = \frac{1}{S_0^{-1}(\omega) + \tilde{t} \alpha({\bf k}) V(\omega)}        \label{Spropag}
\end{equation}
where the on-site Green function $S_0$ is given by
\begin{equation}
S_0^{-1}(\omega) = {\bf I}^{-1}\left(\begin{array}{cc}
(\omega +\mu) I_1 + \tilde{t}e - Dm_0(\omega) & ~-\tilde{t}e + Dm_0(\omega) \\
 ~~~-\tilde{t}e + Dm_0(\omega)~~~ & (\omega +\mu) I_1 + \tilde{t}e - Dm_0(\omega)\end{array}
\right){\bf I}^{-1}, \label{S0propag}
\end{equation}
and the nearest-neighbor contribution is 
\begin{equation}
V(\omega) = \left(\begin{array}{cc}
1 + [\tilde{p} - \tilde{t}^{-1} Dm_1(\omega)]I_1^{-2}~~ & 1 - [\tilde{p} - \tilde{t}^{-1} Dm_1(\omega)]I_1^{-1}I_2^{-1} \\
1 - [\tilde{p} - \tilde{t}^{-1} Dm_1(\omega)]I_1^{-1}I_2^{-1} & 1 + [\tilde{p} - \tilde{t}^{-1} Dm_1(\omega)]I_1^{-2}~~
\end{array}
\right).           \label{Vpropag}
\end{equation}

\subsection{The two-site problem: Level operators and resolvents.}

At this stage, solving our problem entails a calculation of the functions $Dm_0(\omega)$, $Dm_1(\omega)$ and the parameter $\tilde{p}$. To this end, we express these quantities\cite{mm} in terms of correlation functions for the level operators associated with a two-site problem. Let us introduce first the single site level operators $B(i)$, $F_{\sigma}(i)$ and $D(i)$  which annihilate empty, singly occupied (with spin $\sigma$) and doubly occupied states, respectively. In terms of these operators, the original Hubbard operators can be written as $\xi_{\sigma} = B^{\dagger}F_{\sigma}$ and $\eta_{\sigma} = \sigma F_{-\sigma}^{\dagger}D$. 
As the system can be at a given time in one of the possible four states, the level operators satisfy the condition
\begin{equation}
Q(i) \equiv B^{\dagger}(i)B(i) + \sum_{\sigma} F_{\sigma}^{\dagger}(i) F_{\sigma}(i) + D^{\dagger}(i)D(i) = 1.         \label{Qcond}
\end{equation}
This restriction can be introduced by adding a Lagrange multiplier
term of the form, $\varepsilon_0 \sum_iQ(i)$, to the original Hamiltonian.
The level operators for the two-site states, $\Phi_n$, are obtained by taking
 all the combinations of single-site level operators. It is convenient to use the following symmetric or antisymmetric combinations corresponding to eight fermionic-type states,
j\begin{eqnarray}
FB_S^{\sigma} &\equiv& \frac{1}{\sqrt{2}}(F_{\sigma}(x)B(x') + B(x)F_{\sigma}(x')), \nonumber \\
FD_S^{\sigma} &\equiv& \frac{1}{\sqrt{2}}(F_{\sigma}(x)D(x') + D(x)F_{\sigma}(x')),   \nonumber \\
FB_A^{\sigma} &\equiv& \frac{1}{\sqrt{2}}(F_{\sigma}(x)B(x') - B(x)F_{\sigma}(x')), \nonumber \\
FD_A^{\sigma} &\equiv& \frac{1}{\sqrt{2}}(F_{\sigma}(x)D(x') - D(x)F_{\sigma}(x')) ,        \label{phibasis1}
\end{eqnarray}
and eight bosonic-type states,
\begin{eqnarray}
BB &\equiv& B(x)B(x'),~~~~ DD \equiv D(x)D(x'), \nonumber\\
FF^{\sigma} &\equiv& F_{\sigma}(x)F_{\sigma}(x'),\nonumber\\
FF_S &\equiv& \frac{1}{\sqrt{2}}(F_{\uparrow}(x)F_{\downarrow}(x') + F_{\downarrow}(x)F_{\uparrow}(x')) ,\nonumber\\
FF_A &\equiv& \frac{1}{\sqrt{2}}(F_{\uparrow}(x)F_{\downarrow}(x') - F_{\downarrow}(x)F_{\uparrow}(x')) ,\nonumber\\
DB_S &\equiv& \frac{1}{\sqrt{2}}(D(x)B(x') + B(x)D(x')), \nonumber\\
DB_A &\equiv& \frac{1}{\sqrt{2}}(D(x)B(x') - B(x)D(x')),       \label{phibasis2}
\end{eqnarray}
where $x$ and $x'$ are the positions of the two-sites and the spin index, $\sigma$, takes two values, ($+1 = \uparrow$) and ($-1 = \downarrow$). Note that $FF^{\sigma}$ and $FF_S$ correspond to the spin-triplet states, while $FF_A$ corresponds to the  spin-singlet state. 

We are interested in solving a two-site problem for a cluster embedded in a reservoir, constituted by the rest of the system. Formally,
the total Hamiltonian can be divided into three parts\cite{mm} describing the two-site sub-system, $H_0$, the reservoir, $H_R$, and their interaction, $H_{0R}$,
\begin{equation}
H = H_0 + H_R + H_{0R}.
\end{equation}
To describe the properties of the two-site system, we introduce the resolvent 
\begin{equation}
R_{nm}(t-t') = \theta(t-t')\frac{Tr_R\left[ \langle 0 | \Phi_n(t)\Phi_m(t')^{\dagger}|0\rangle e^{-\beta H_R}\right]}{Tr_R \left[e^{-\beta H_R}\right]}, 
\end{equation}
where the trace is over the degrees of freedom of the reservoir, $|0\rangle$ denotes the vacuum for the two-site problem, and, as usual, $\beta = 1/k_BT$. Note that $\Phi_n|0\rangle = 0$ and , consequently, $H_{0R}|0\rangle = 0$.
The Fourier transform of the resolvent can be expressed using the spectral function, $\sigma_{nm}(\omega) = -\frac{1}{\pi} {\rm Im}R_{nm}$,
\begin{equation}
R_{nm}(\omega) = \int dx\frac{\sigma_{nm}(\omega)}{\omega - x + i\delta}.
\end{equation}
We also introduce the auxiliary function $\bar{\sigma}_{nm}(\omega) = e^{-\beta \omega}\sigma_{nm}(\omega)$. 
Once we know the resolvents, we can express any average of operators of the type $\Phi_{nm} = \Phi_m^{\dagger}\Phi_n$ as
\begin{equation}
\langle\Phi_{nm}\rangle = \frac{1}{Z}\int d\omega~\bar{\sigma}_{nm}(\omega) 
\end{equation}
with $Z = \sum_n\int d\omega~\bar{\sigma}_{nn}(\omega)$.

A formal solution of the resolvents\cite{mm} can be obtained using the equation of motion method. We can write
\begin{equation}
R_{nm}(\omega) = \left(\frac{1}{\omega - E - \Sigma(\omega)}\right)_{nm}, \label{RnmEq}
\end{equation}
where the energy matrix $E$ is determined by the levels of an isolated two-site system and the self-energy $\Sigma$ is a measure of the effects of the reservoir. Explicitly, we have
\begin{equation}
E_{nm} = \langle0|i\frac{\partial}{\partial t} \Phi_n(t) \Phi_m^{\dagger}(t)|0\rangle_R 
\end{equation}
and 
\begin{equation}
\Sigma_{nm}(\omega) = F.T. \langle0|\theta(t-t') \delta J_{\Phi_n}(t) \delta J_{\Phi_n}^{\dagger}(t')|0\rangle_{RI}, \label{selfE}
\end{equation}
where `R' indicates that the trace over the reservoir degrees of freedom has been taken, `I' indicates the irreducible part, and $\delta J_{\Phi_n}(t) = i\frac{\partial}{\partial t} \Phi_n(t) - \sum_m E_{nm}\Phi_m(t)$.
To proceed, we determine the equations of motion for the level operators. It is convenient to write first the equations for the single-site operators,
\begin{eqnarray}
i\frac{\partial}{\partial t} B~ &=& \varepsilon_0 B - \tilde{t}\sum_{\sigma} c_{\sigma}^{\dagger\alpha}F_{\sigma}, \nonumber\\
i\frac{\partial}{\partial t} F_{\sigma} &=& (\varepsilon_0 -\mu)F_{\sigma} =\tilde{t} B c_{\sigma}^{\alpha} - \tilde{t}\sigma c_{-\sigma}^{\dagger\alpha} D , \label{singeq}\\
i\frac{\partial}{\partial t} D~ &=& (\varepsilon_0 -2\mu +U)D - \tilde{t}\sum_{\sigma}\sigma F_{-\sigma} c_{\sigma}^{\alpha}, \nonumber
\end{eqnarray}
where the arbitrary reference energy $\varepsilon_0$ will be set to $\varepsilon_0 = -\mu$.  
In fact, the lattice and the two-site cluster
had differing chemical potentials.  However, equilibrium
between the two-site system and the lattice requires that both 
have the same chemical potential.  
The equations of motion for the two-site level operators can be determined directly using the equations of motion, Eq. (\ref{singeq}) and Eqs. (\ref{phibasis1}) and (\ref{phibasis2}). Explicitly, these equations are given in Appendix A. From the equations of motion we can extract the energies $E_{nm}$ for the resolvents. Selecting the terms that do not contain $c_{\sigma}^{\bar{\alpha}}$-type operators, that is, the terms that do not depend on the degrees of freedom of the reservoir, we obtain:
\begin{eqnarray}
E_{FB_S} &=& 2\varepsilon_0 -\mu-\frac{\tilde{t}}{2d}, ~~~ 
E_{FD_S} = 2\varepsilon_0 -3\mu +U + \frac{\tilde{t}}{2d}, \nonumber \\
E_{FB_A} &=& 2\varepsilon_0 -\mu+\frac{\tilde{t}}{2d}, ~~~ 
E_{FD_A} = 2\varepsilon_0 -3\mu +U - \frac{\tilde{t}}{2d}, \nonumber \\
E_{BB_{~}} &=& 2\varepsilon_0, ~~~~~~~~~~~~~~~~
E_{DD_{~}} = 2(\varepsilon_0 -2\mu +U)  \label{energnm} \\
E_{FF_S} &=& E_{FF^{\sigma}} = E_{FF_A} = 2(\varepsilon_0-\mu), \nonumber \\
E_{DB_S} &=& E_{DB_A} = 2\varepsilon_0 - 2\mu +U, ~~E_{DB_S~FF_A} = E_{FF_A~DB_S} = \frac{\tilde{t}}{d}.\nonumber 
\end{eqnarray} 
Employing the standard non-crossing approximation\cite{C4ncapp} and using the time derivatives of the $\Phi_n$ operators from Appendix A, we compute the self-energies $\Sigma_{nm}$, Eq. (\ref{selfE}), of the resolvents.  The expressions for the self-energies are given in Appendix B.

It is known\cite{C4NCA} that inter-site spin fluctuations, which are ignored in the non-crossing approximation, are in fact important and cannot be neglected at energy scales on the order of $t^2/U$.  Since our attempt is to develop a top-down approach
in which on-site physics as well as nearest-neighbour correlations
are described accurately, we must include spin fluctuations.  To overcome this problem, we include the effects of spin fluctuations as a higher order correction to the self-energies of the resolvents.  Physically, we can understand this correction by observing that the solution for singlet and triplet states, $FF_A$ and $FF_S$, is sharply peaked at energies separated by
\begin{equation}
J = \int_{-\infty}^{\infty} \omega (\sigma_{FF_S} - \sigma_{FF_A}) ~d\omega. 
\label{Jequation}
\end{equation}
In the strong coupling limit, $J$ is of order $t^2 /U$ and, consequently, singlet-triplet mixing cannot be ignored.  We consider $J$ to be the coupling constant of an effective antiferromagnetic interaction which is responsible for the spin fluctuations. The corrections to the self-energies given by spin fluctuations are given in Appendix C. The energies from Eq. (\ref{energnm}) , together with the self-energies given in Appendix B and the corrections from Appendix C constitute the equations necessary for the evaluation of the two-site resolvents.

\subsection{Self-consistent procedure}

\begin{figure}
\centering
\includegraphics[height=7cm]{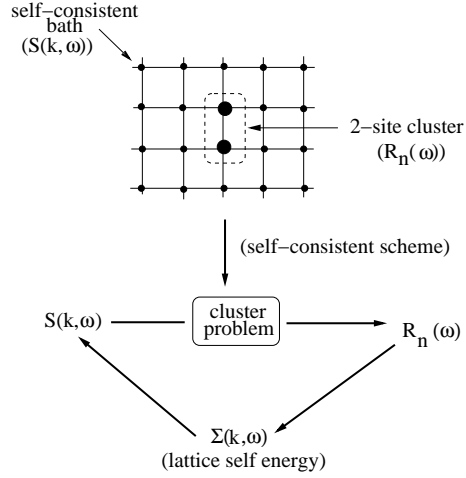}
\caption{Self-consistent scheme for the computation of the self-energy.}
\label{scheme}
\end{figure}

The goal of  introducing the two-site resolvents is to express the dynamical corrections $Dm_0$ and $Dm_1$, as well as the parameter $\tilde{p}$ in terms of quantities associated with the two-site problem as depicted in
 Fig. (\ref{scheme}). 
To this end, let us introduce the quantity $Z_{\phi_n} = Z\langle \Phi_n^{\dagger}\Phi_n\rangle$ by
\begin{equation}
Z_{\Phi_n} = \int d\omega~\bar{\sigma}_{\Phi_n}(\omega).  \label{zphi}
\end{equation}
The two-site occupation numbers for each of the 16 states can be written directly in terms of $Z_{\Phi_n}$. For example, for the singlet and triplet states, we have $n_{FF_A} = Z_{FF_A} / Z$ and $n_{FF_S} = 3Z_{FF_S} / Z$, respectively. Single-site occupation numbers can also be expressed in terms of $Z_{\Phi_n}$  using Eq. (\ref{Qcond}) and writing the identity operator on a neighboring site as ${\bf 1}' = Q(x')$. For example, 
\begin{equation}
D^{\dagger}D \equiv D^{\dagger}D {\bf 1}' = (DB')^{\dagger} DB' + \sum_{\sigma} (DF_{\sigma}')^{\dagger}DF_{\sigma}' + (DD')^{\dagger}DD',
\end{equation}
and as a result
\begin{equation}
n_D \equiv \langle D^{\dagger}D \rangle = Z^{-1}\left[Z_{FD_S} +Z_{FD_A} + \frac{1}{2}Z_{DB_S} + \frac{1}{2}Z_{DB_A} + Z_{DD}\right].
\end{equation}
Similarly, we obtain for the average single occupancy, $\sum_{\sigma}\langle F_{\sigma}^{\dagger} F_{\sigma} \rangle$,
\begin{equation}
n_F = Z^{-1}\left[3Z_{FF_S} +Z_{FF_A} + Z_{FB_S} + Z_{FB_A} + Z_{FD_S} + Z_{FD_A}\right].
\end{equation}
The electron filling is given by the sum of single occupancy and (twice) the double occupancy,
\begin{equation}
n = n_F + 2n_D .
\label{SelfCn}
\end{equation}
As all the terms in $\tilde{p}$ can be expressed in terms of products of two-site operators, it is straightforward to write this parameter as
\begin{equation}
\tilde{p} = Z^{-1}\left[ Z_{FD_S} +  Z_{FD_A}  + Z_{DD} +\frac{3}{2}Z_{FF_S} - \frac{1}{2}Z_{FF_A} - \frac{1}{2}Z_{DB_S} + \frac{1}{2}Z_{DB_A}\right] -\frac{n^2}{4}.       \label{pCorrdy}
\end{equation}
This represents the self-consistency condition for the parameter $\tilde{p}$ and obviates the need to employ a decoupling
scheme 
required
in the static approximation\cite{hubbard,compo}.

The next step is to determine the dynamical corrections. To this end, we
 express\cite{mm} the 'currents' $\delta J_{\sigma}$ in terms of two-site level operators. From Eq. (\ref{delphi}) and $x_i \rightarrow x$ and $x_I^{\alpha} \rightarrow x'$, we 
find that
\begin{eqnarray}
\delta J_{\sigma} &=& -\tilde{t}\sum_{m,n} a_{nm} \Phi_n^{\dagger}\Phi_m, \nonumber\\
\delta J_{\sigma}^{\prime} &=& -\tilde{t}\sum_{m,n} a_{nm}^{\prime} \Phi_n^{\dagger}\Phi_m,                      \label{Jexpan}
\end{eqnarray}
where $\Phi_n$ represents the complete set of two-sites level operators and the coefficients $a_{nm}$ are given in Appendix D. 
As $\delta J^{\prime}$ is obtained from $\delta J$ by exchanging the positions $x$ and $x'$, the coefficients $a_{nm}^{\prime}$ will be identical with $a_{nm}$ up to a sign that depends on the symmetry properties of the states '$\Phi_m$' and '$\Phi_n$' under the exchange of $x$ and $x'$. Let us denote by ${\cal E}$ the operator that produces the exchange,
\begin{equation}
{\cal E}O(x,x') = O(x',x),
\end{equation}
where $O(x,x')$ is an arbitrary operator defined on the two-site cluster.
The symmetry properties of the level operators are given by
\begin{eqnarray}
{\cal E} FB_S &=& ~FB_S, ~~~{\cal E} FB_A = -FB_A,  ~~~{\cal E} FD_S = FD_S, ~~~{\cal E} FD_A = -FD_A,  \nonumber\\
{\cal E} FF_S &=& -FF_S, ~~~{\cal E} FF^{\sigma} = -FF^{\sigma}, ~~~{\cal E} FF_A = FF_A,    \\
 {\cal E} DB_S &=& ~DB_S, ~~~{\cal E} DB_A = -DB_A,  ~~~{\cal E} BB = BB, ~~~{\cal E} DD = DD.  \nonumber
\end{eqnarray}
Consequently, $a_{nm}^{\prime} = a_{nm}$ if $\Phi_n$ and $\Phi_m$ have the same symmetry and $a_{nm}^{\prime} = -a_{nm}$ otherwise.
The self-energy contributions can be calculated only approximately. We will use the non-crossing approximation\cite{C4ncapp} which has been proven to be effective in solving problems with local correlations. 
Defining $\Phi_{mn} = \Phi_n^{\dagger}\Phi_m$, we have
\begin{eqnarray}
&Tr_R&\langle 0|\Phi_{mn}(t)\Phi_{m'n'}^{\dagger}(t')e^{-\beta H}| 0 \rangle \\
&=& Tr_R\left[ \langle 0|\Phi_m(t)\Phi_{m'}^{\dagger}(t')| 0 \rangle\langle 0|\Phi_{n'}(t') e^{-\beta H} \Phi_n^{\dagger} (t)| 0 \rangle\right] \nonumber \\
&\approx& Tr_R\langle 0|\Phi_m(t)\Phi_{m'}^{\dagger}(t')| 0 \rangle Tr_R\langle 0|\Phi_{n'}(t')\Phi_{n}^{\dagger}(t+i\beta)| 0 \rangle. \nonumber
\end{eqnarray}
Consequently, we obtain
\begin{equation}
\langle \Phi_{mn}(t)\Phi_{m'n'}^{\dagger}(t')\rangle \approx
\frac{1}{Z}\int d\omega ~e^{-i\omega(t-t')}\int dx~\sigma_{mm'}(\omega+x)\bar{\sigma}_{n'n}(x), \label{NCAeq}
\end{equation}
which is the formulation of the non-crossing approximation that we will use systematically in our calculations.
Introducing the expansion Eq. (\ref{Jexpan}) in the expression for the dynamical corrections, Eq. (\ref{DmEq}), and using the non-crossing approximation formula( Eq. (\ref{NCAeq})), we obtain
\begin{equation}
Dm_0(\omega) = \frac{1}{2dZ} \int dx~dx'~\sum_{n,m,n',m'} a_{nm}a^{\star}_{n'm'} \frac{\sigma_{mm'}(x)\bar{\sigma}_{n'n}(x') + \bar{\sigma}_{mm'}(x)\sigma_{n'n}(x')}{\omega -x + x' + i\delta}
\end{equation}
and 
\begin{equation}
Dm_1(\omega) = \frac{1}{2dZ} \int dx~dx'~\sum_{n,m,n',m'} a_{nm}a^{\prime\star}_{n'm'} \frac{\sigma_{mm'}(x)\bar{\sigma}_{n'n}(x') + \bar{\sigma}_{mm'}(x)\sigma_{n'n}(x')}{\omega -x + x' + i\delta}
\end{equation}
Detailed expressions for $Dm_0$ and $Dm_1$ are given in Appendix D.  

This concludes the process of writing the components of the Green function 
Eq. (\ref{greenEq}) that cannot be expressed in terms of the Hubbard basis, in terms of spectral functions for the two-site system. However, for a fully self-consistent calculation, we need to determine the reservoir spectral functions $\rho_{\pm}(\omega)$ that enter into the formulas for the self-energies of the resolvents (see Appendix B). The reservoir consists of the full system from which the two sites $x$ and $x'$ have been excluded. Our goal is to determine the Green functions $\bar{G} = \langle\langle c_{\sigma}^{\bar{\alpha}},c_{\sigma}^{\dagger\bar{\alpha}}\rangle \rangle$ and  $\bar{G}' = \langle\langle c_{\sigma}^{\bar{\alpha}},c_{\sigma}^{\dagger\bar{\alpha}'}\rangle \rangle$. We introduce the two-site full propagator $[S] = \langle\langle \Psi, \Psi^{\dagger}\rangle \rangle$ with  $\Psi^{\dagger} = (\psi^{\dagger}(x) \psi^{\dagger}(x'))$ and the irreducible propagator 
\begin{equation}
[\bar{S}] = \left(\begin{array}{cc}
\bar{S} & \bar{S}' \\
\bar{S}' & \bar{S}
\end{array}
\right).        
\end{equation}
where $\bar{S} = \langle\langle \psi^{\bar{\alpha}}, \psi^{\dagger\bar{\alpha}}\rangle \rangle$ and $\bar{S}' = \langle\langle \psi^{\bar{\alpha}}, \psi^{\dagger\bar{\alpha}'}\rangle \rangle$.
From standard scattering theory, we have that 
\begin{equation}
[S] = [S_0] + [S_0][V][\bar{S}][V][S]  
\end{equation}
where 
\begin{equation}
[V] = \left(\begin{array}{cc}
V & 0 \\
0 & V
\end{array}
\right)       
\end{equation}
and 
\begin{equation}
[S_0]^{-1} = \left(\begin{array}{cc}
S_0^{-1} & tV \\
tV & S_0^{-1}
\end{array}
\right).        
\end{equation}
The solution of these equations is
\begin{eqnarray}
\bar{S}  &=& V^{-1}\left(S_0^{-1} - (S-S'S^{-1}S')^{-1}\right)V^{-1}, \nonumber\\
\bar{S}' &=& V^{-1}\left(tV + S^{-1}S'(S-S'S^{-1}S')^{-1}\right)V^{-1}.
\end{eqnarray} 
The irreducible Green functions used in the evaluation of the self-energies of the resolvents will be given by
\begin{eqnarray}
\bar{G} &=& \bar{S}_{11} +  \bar{S}_{12} + \bar{S}_{21} + \bar{S}_{22}, \nonumber\\
\bar{G}' &=& \bar{S}'_{11} +  \bar{S}'_{12} + \bar{S}'_{21} + \bar{S}'_{22}, 
\end{eqnarray} 
and the corresponding spectral functions are
\begin{eqnarray}
\rho_+(\omega) &=& -\frac{2}{\pi} {\rm Im}(\bar{G}(\omega) + \bar{G}'(\omega)) ,\nonumber\\
\rho_-(\omega) &=& -\frac{2}{\pi} {\rm Im}(\bar{G}(\omega) - \bar{G}'(\omega)). \label{irred}
\end{eqnarray}

\section{Results}

We now have all the ingredients necessary for the implementation of the self-consistent procedure shown in Fig. (\ref{scheme}).  Starting with an initial guess for the spectral functions $\rho_{\pm}$  describing the properties of the environment, we solve the two-site problem iteratively using the expression Eq. (\ref{RnmEq}) for the resolvents, the energies Eq.~(\ref{energnm}) and the self-energies from Appendix B with the spin fluctuation corrections from Appendix C. While in principle the cluster and the 
lattice can have different chemical potentials as in the previous
work on the two-site cluster\cite{mm}, we have used the more physical
restraint that both the cluster and the lattice must have the same 
chemical potential.  However, as a result of Eq. (\ref{SelfCn}),
the filling in the cluster and the lattice will be different.   By symmetry,
however, at half-filling both the cluster and the lattice will have the same
filling.   All the frequency-dependent functions are discretized on a grid of $N = 8192$ points from $\omega_{min} = -20 t$ to $\omega_{max} = 20 t$. To increase the computational speed, we performed all the convolutions involved in the calculation of the self-energies using a fast Fourier transform algorithm. The procedure converges for temperatures above $T = 0.02t$ at finite doping and $T = 0.08t$ at half-filling, although convergence problems occurred below $T=0.1t$. Once we computed the resolvents, the mean-field parameter $\tilde{p}$ can be determined using Eq. (\ref{pCorrdy}), as well as the dynamical corrections $Dm_0$ and $Dm_1$ from Appendix D.  The mean-field parameter $e = \langle\xi^{\alpha}\xi^{\dagger}\rangle - \langle\eta^{\alpha}\eta^{\dagger}\rangle$ can be expressed in terms
of the Green function using the general self-consistency condition,
\beq\label{corr1}
\langle\psi_m(i)\psi^\dagger_n(j)\rangle&=&\frac{\Omega}{(2\pi)^2}\int d^2k d\omega
e^{i\vec k\cdot(\vec r_i-\vec r_j)}(1-f(\omega))\nonumber\\
&&\times\left(\frac{-1}{\pi}\right)
{\rm Im} S_{mn}(\bf k,\omega). \nonumber\\
\eeq 
Within the grand-canonical ensemble, the chemical potential, $\mu$, is determined
by the self-consistent solution to
\beq\label{mutot}
n=2\left(\langle\xi\xi^\dagger+2\xi\eta^\dagger+\eta\eta^\dagger\rangle\right)
\eeq
We imposed the constraint that the chemical from Eq. (\ref{mutot}) also equal that for the cluster. 
We then determined the full Green function $S({\bf k}, \omega)$  using Eqs. (\ref{Spropag}-\ref{Vpropag}).  
Next, new  spectral functions $\rho_{\pm}$ are determined using Eq. (\ref{irred}) and the whole procedure (see Fig. (\ref{scheme}) is repeated until full convergence is reached.

\subsection{Spectral Function at Half-filling: Mott Insulator} 

Before we analyze the doped case, we first review\cite{stanescu1} the properties of 
the charge vacuum that determines the insulating behavior at half-filling.
To reiterate, there are two distinct
routes to the insulating
state at half-filling.  For $U \gg t$, the charge and spin degrees of freedom are decoupled and the system is an insulator for temperatures smaller than $T_0 \sim U$. The spins are coupled due to the super-exchange interaction, $|J| \approx 4t^2/U$. It is this spin exchange interaction
that gives rise to local antiferromagnetic fluctuations and eventually ordering at $T=0$.  This is the 
antiferromagnetic Mott insulating state. In general, in the weak coupling regime, a metal-insulator transition occurs as a consequence of the Brillouin zone folding generated by magnetic or charge ordering and the corresponding gap is essentially related to antiferromagnetism or some
 type of charge density wave.  This type of transition is referred to as a Slater transition and the corresponding insulating state should not be confused with
the Mott insulator. Such a regime can be successfully described by conventional many-body approaches.
 
\begin{figure}
\centering
\includegraphics[height=6.0cm]{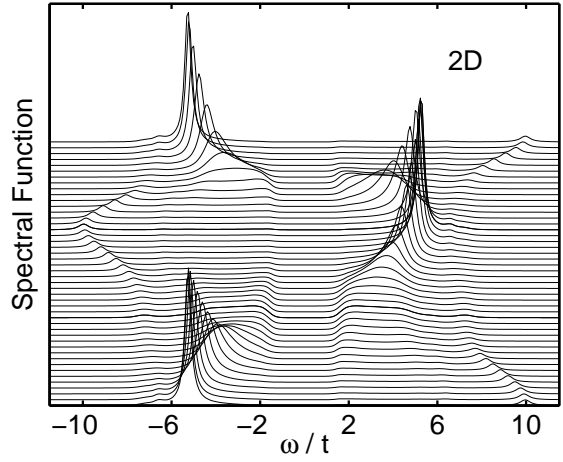}
\caption{Momentum and energy dependence of the electron spectral function for a half-filled 2D system with $U=8t$ and $T=0.15t$. From top to bottom, the curves correspond to $(k_x,k_y) = (0,0) \rightarrow (\pi,\pi) \rightarrow (\pi,0) \rightarrow (0,0)$.} 
\label{fig4.1a}
\end{figure}

\begin{figure}
\centering
\includegraphics[height=5.5cm]{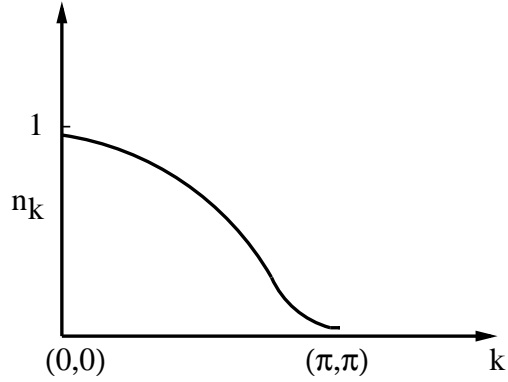}
\caption{Occupancy of each momentum state with $k_x = k_y$ for the spectral function shown in
Fig.(\protect\ref{fig4.1a}).  The occupancy is peaked at $(0,0)$ and decreases 
continuously to a minimum value 
at $(\pi,\pi)$ without the discontinuity indicative of a Fermi liquid.}
\label{nk}
\end{figure}

Shown in Fig. (\ref{fig4.1a}) is the total electron spectral function,
\beq
A(\vec k,\omega)=-\frac{1}{\pi} {\rm Im}\left\{S_{11}({\bf k}, \omega) + 2S_{12}({\bf k}, \omega) + S_{22}({\bf k}, \omega)\right\}
\eeq
 for the 2D half-filled
Hubbard model with $U=8t$ and $T=0.15t$.
\noindent 
Clearly visible are the upper and lower Hubbard bands with an
energy gap of order $U$ and flatness of the band near the
$(\pi,0)$ point.
The chemical potential ($\omega=0$)
lies in the middle of the gap and hence the system is an insulator. However,
no symmetry is broken as is evident because the periodicity 
is $2\pi$ rather than $\pi$ as would be the case if the Brillouin zone
had doubled. In addition, spin and charge are not fractionalized\cite{sl5}.
 The insulating behavior arises because the charge gap has splintered the spectral weight of each $\vec k$-state into `bonding' and `antibonding'
pieces. Consequently, there is a fundamental breakdown of what is meant by
an electronic state.  In fact, the electronic states themselves
have fractionalized.  To make contact with the
real-space picture shown in Fig. (\ref{fig1}), we note that the PES and IPES spectra are determined by the spectral weight in the lower and upper Hubbard bands,
respectively.  In general, the upper and lower Hubbard bands carry total
 spectral
weight $n/2$ and $1-n/2$, respectively which of course reduces to $1/2$ at half-filling.  However, $A(\vec k,\omega)$ is strongly momentum
dependent as illustrated in Fig. (\ref{fig4.1a}).  For the lower Hubbard band, the maximum in the spectral weight is peaked at $(0,0)$ and decreases as the $(\pi,0)$ point is reached and becomes vanishingly small at $(\pi,\pi)$.  In fact, the states at $(\pi,\pi)$ in the lower Hubbard band carry almost none of the spectral weight.  However, the decrease in the occupancy of each $\vec k$-state is a continuous function, as depicted in Fig. (\ref{nk}), rather than a discontinuous one as would be the case in a Fermi 
liquid with a well-defined Fermi surface. In the strict sense, the discontinuity in $n_{\vec k}$ in a Fermi liquid occurs at $T=0$.  Although we cannot reach $T=0$ in our approach, we find no indication that a discontinuity develops in $n_{|vec k}$ as the temperature decreases.  In fact, the continuous behavior
we have obtained here is consistent with the exact result\cite{eskes}
\beq
n_{\vec k}=\frac12+2\frac{\epsilon_{\vec k}}{U}\langle \vec S_i\cdot\vec S_{i+\vec \delta}-\frac14\rangle
\eeq
for the occupancy in each $\vec k-$state projected into the LHB
of a half-filled Hubbard model. 
Note all the $\vec k-$ dependence is determined by the single
particle energies,
\beq
\epsilon_{\vec k}=-t\sum_{\vec \delta} e^{i\vec k\cdot\left(\vec R_i-\vec R_{i+\delta}\right)}
\eeq
For a paramagnetic state, the term in the angle brackets is exactly, $-1/4$.  Hence, $n_{\vec k}$ is continuously decreasing function from $(0,0)$ to $(\pi,\pi)$ as found here for $U=8t$.

\noindent
\noindent
Although $n_{\vec k}<1$, each state satisfies the sum rule,
\beq
\int_{-\infty}^\infty A(\vec k,\omega)d\omega=1.
\eeq
In a band insulator the upper cutoff on the energy is simply the chemical potential.   This does not mean that in a band insulator or in a Fermi liquid,
a broad incoherent background cannot be present which extends to high energies. In fact, in a Fermi liquid, $n_{\vec k}$ can be less than unity. 
However, in Fermi liquids, a coherent quasi-particle peak always exists regardless of the momentum.  Hence, the criterion for occupancy of a single particle
state is simply whether or not the coherent peak lies above or below the chemical potential. In a Mott insulator not only is the spectral weight split over an energy scale of $U$ but there is an absence of coherent quasi-particles as evidenced by the broad spectral features. 
 Hence, there is no sharp criterion for 
unit occupancy of a single particle state.  The broadness of the spectral features stems from the local correlations on neighbouring sites not the Mott gap itself.  Without the dynamical corrections,
the spectral function would simply be a sum of $\delta-$function peaks at the lower and upper Hubbard
bands.  

The bifurcation of the spectral weight of each $\vec k-$state above and below
the charge gap can be modeled as follows.  Consider for
example $\gamma^\dagger_{\vec k\sigma}=u_{\vec k}\xi^\dagger_{\vec k\sigma}+v_{\vec k}\eta^\dagger_{\vec k\sigma}$, with coefficients
$u_{\vec k}$ and $v_{\vec k}$ are determined by the projection of the spectral
function onto the lower and upper Hubbard bands, respectively.
Hence, the antisymmetrized state formed from such excitations
\beq\label{MI}
|\rm MI\rangle=\sum_{\rm P}(-1)^P\prod_{\vec k \in \rm FBZ}
\gamma^\dagger_{\vec k\uparrow}\gamma^\dagger_{\vec k\downarrow}|0\rangle
\eeq
is a candidate for describing the elusive paramagnetic Mott insulator.  Provided
magnetic  frustration is present so that ordering is pre-empted, Eq. (\ref{MI}) should be the $T=0$ Mott
insulating state.    
The sum over all permutations, $P$, is necessary as
the $\gamma_{\vec k\sigma}$ operators obey the
non-fermionic commutation relations,
\beq\label{comm}
\{\gamma^\dagger_{\vec k\uparrow}, \gamma^\dagger_{\vec q\downarrow}\}=
\sum_i e^{-i(k+q)\cdot r_i} (u_kv_q-u_qv_k)c^\dagger_{i\uparrow}
c^\dagger_{i\downarrow}
\eeq
and $\{\gamma^\dagger_{\vec k\sigma}, \gamma^\dagger_{\vec q\sigma}\}=
\{\gamma^\dagger_{\vec k\sigma}, \gamma_{\vec q\sigma'}\}=0$. 
When $u_k=v_k$, 
$\gamma^\dagger_{\vec k\uparrow}\gamma^\dagger_{\vec k\downarrow}|0\rangle$
generates the completely doubly occupied state.  However,
because the right-hand side
of the anticommutation relation in Eq. (\ref{comm}) is identically zero
for $u_k=v_k$, the zero state results upon summation over all permutations.
Consequently, Eq. (\ref{MI}) completely projects out the fully
doubly occupied state.

\subsubsection{Singlet Formation: Local Antiferromagnetism}

\begin{figure}
\centering
\includegraphics[height=5.5cm]{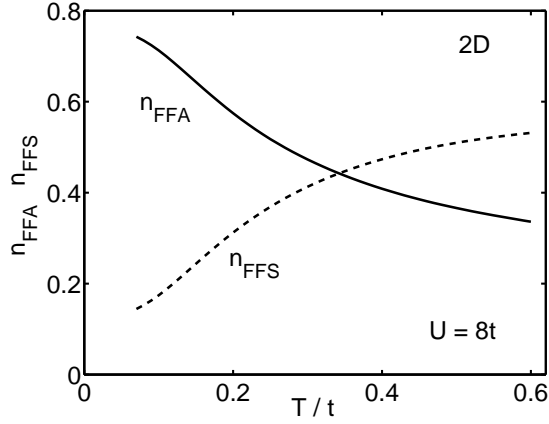}
\caption{Singlet ($n_{FF_A}$) and triplet ($n_{FF_S}$) occupation numbers as a function of temperature  for $U=8t$ in two dimensions. 
 The fact that $ n_{FF_A} > n_{FF_S}$ as $T\rightarrow 0$ is consistent with
  local antiferromagnetic order.}
\label{fig4.8a}
\end{figure}

A crucial test of the correctness of the method we have used here
is whether or not
short-range antiferromagnetic correlations are present 
at low temperatures. Such correlations do not signal that long-range magnetic order obtains at $T=0$ but rather that the ground state at $T=0$ is a liquid of nearest-neighbour singlet states as in the resonating valence bond (RVB) state proposed by Anderson\cite{anderson}.  We are able
with our two-site formalism to probe the existence of local magnetic order
by computing 
the nearest-neighbour singlet and triplet occupation numbers, $n_{FF_A} = Z_{FF_A}/Z$ and $n_{FF_S} = 3Z_{FF_S}/Z$, respectively.
From Fig. (\ref{fig4.8a}) we find that, at high temperatures,
triplet excitations dominate.  However, this trend is reversed below some
temperature and the singlet occupancy becomes of order unity.
Hence, the low-temperature properties of the insulating state we have computed
here are consistent with a liquid of nearest-neighbour singlet states as in the RVB state.  In fact, at $T=0$ the liquid state we have found here persists
because we have imposed a paramagnetic solution.  

\begin{figure}
\centering
\includegraphics[height=6.0cm]{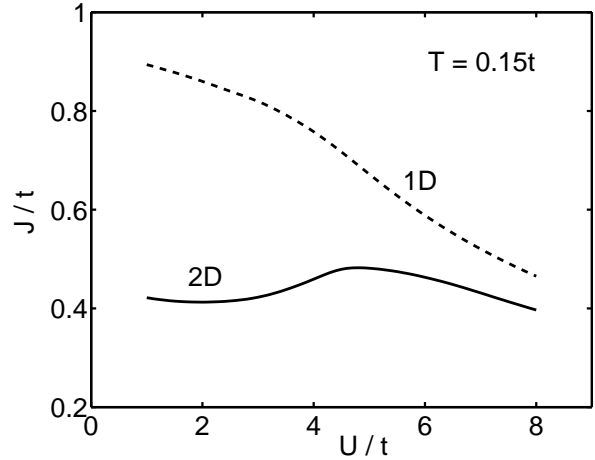}
\caption{Effective exchange interaction coupling constant $J$ as a function of $U/t$ for $T=0.15t$.}
\label{fig4.9a}
\end{figure}

A final diagnostic of the insulating state we have found here is the behavior
of the effective exchange interaction as a function of $U$.  In the Mott state,
a super-exchange interaction is self-generated which should scale as $1/U$.
It is this exchange interaction that sets the scale for the Ne\'el temperature. Using Eq. (\ref{Jequation}), we computed
the effective exchange interaction shown in Fig. (\ref{fig4.9a}) for both $1D$ and
$2D$.  Note first that $J$ is always positive as a consequence of the fact
that the singlet state is lower in energy than the triplet. This is a further
indication of antiferromagnetic correlations in the ground state.
As expected, $J$ is well approximated
by $4t^2/U$ in the strong-coupling regime.  However,
as $U$ decreases, deviations from this behavior are observed.  

\subsection{ Doped Mott Insulators}

\subsubsection{Chemical Potential}

\begin{figure}
\centering
\includegraphics[height=7.0cm]{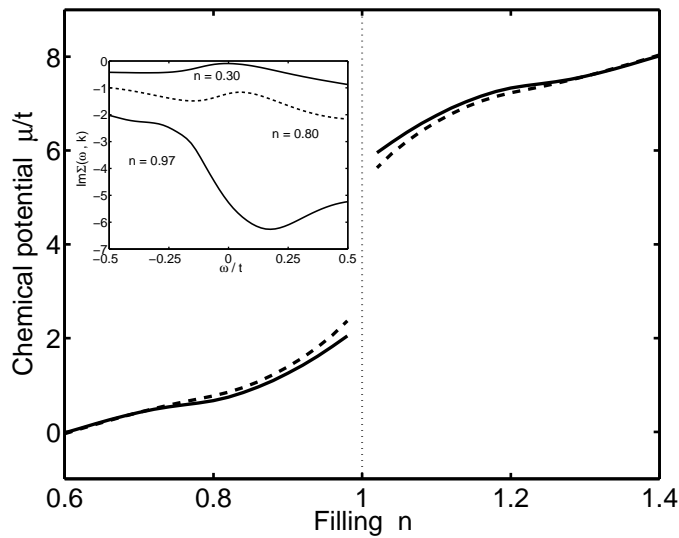}
\caption{Doping dependence of the chemical potential in the 2D Hubbard model
computed using the local cluster approach for $T=0.15t$ (dashed line)
and $T=0.07t$ (solid line).  The inset shows the imaginary
part of the self energy evaluated at a Fermi momentum ($0.3,2.10$) for $n=0.97$,
($0.3,1.84$) for $n= 0.8$ and ($0.3,1.06$) for $n=0.3$.
}
\label{chempot}
\end{figure}

Two scenarios are possible for the doping dependence of the chemical 
potential: 1) the chemical potential remains pinned and mid-gap states are generated by some physical mechanism, or 2) the chemical potential jumps to the top of the lower Hubbard band (LHB)
or the bottom of the upper Hubbard band (UHB) upon hole or electron doping, respectively.  Our results shown in Fig. (\ref{chempot}) demonstrate that the 
chemical potential jumps upon hole or electron doping, indicating an absence 
of mid-gap states.  
The Magnitude of the jump is set by the Mott gap which
is fully developed at $T=0$. 
\noindent
While at some
finite temperature the chemical potential may appear to evolve
smoothly, at $T=0$, the chemical potential should jump discontinuously. 
That is, we find that doping
a Mott insulator leads to a continuous depletion of the spectral weight 
in the first Brillouin zone. Our finding is consistent with the exact result for the 1D Hubbard model\cite{1D} as well as
Quantum Monte-Carlo simulations\cite{qmc,scalapino} in 2D.  
However, in $d=\infty$, the chemical potential exhibits\cite{kotliar} a jump
but one that is smaller than the gap.  Hence, mid-gap states are generated\cite{kotliar}.
 It would appear then that $d=\infty$ is
vastly different from an actual finite-dimensional system and the 2D Hubbard model is quite similar to its 1D counterpart at least as the chemical potential
is concerned. The possible source of the discrepancy is the form of the 
self energy.  A chemical potential jump requires a large imaginary part of the self-energy at the chemical potential, thereby indicating an absence of well-defined quasi-particles.  Mid-gap states are resonance states and hence are reminiscent of the Brinkman/Rice\cite{br} mechanism for the insulator-metal transition
in the doped Mott state.  The inset in Fig. (\ref{chempot}) indicates that $\Im\Sigma$ in the underdoped regime is large and non-zero at the Fermi energy.  Such behavior points to an absence of well-defined quasiparticles.  In the overdoped regime, the characteristic $\omega^2$ dependence appears, indicative of a
Fermi liquid.  Consequently, the method we use here is capable of recovering 
Fermi liquid theory in the overdoped regime. 

Experimentally, whether the chemical potential is pinned or moves
upon doping appears to be cuprate dependent. For example,
in La$_{2-x}$Sr$_x$CuO$_4$\cite{lsco} (LSCO), the 
chemical potential remains pinned roughly at $0.4eV$ above the top of the
LHB, while for  Nd$_{2-x}$Ce$_x$CuO$_4$ (NDCO)\cite{harima1},
Bi$_2$Sr$_2$Ca$_{1-x}$R$_x$Cu$_2$O$_{8+y}$ 
(BSCO)\cite{harima2,hassan,ritveld,tjernberg}, 
and Na-doped Ca$_2$CuO$_2$Cl$_2$\cite{cuclo} (CACLO) the chemical potential jumps upon doping by an amount in accordance with half the Mott gap and scales roughly as $\delta^2$ as obtained here.  Because stripes or macroscopic phase separation require
 the 
chemical potential to be pinned, they have been
invoked\cite{kivelson} to explain the origin of mid-gap states in LSCO.
The pseudogap in the underdoped cuprates has also been attributed\cite{kivelson} to stripes.
However, because $\Delta\mu\ne 0$ for most of the cuprates, for example,
 NDCO, BSCO, and CACLO, if the pseudogap has a universal origin,
stripes are not its cause.  The precise origin of the pseudogap will be discussed extensively in a later section.  

\subsubsection{Spectral Function}

\begin{figure}
\centering
\includegraphics[height=10cm]{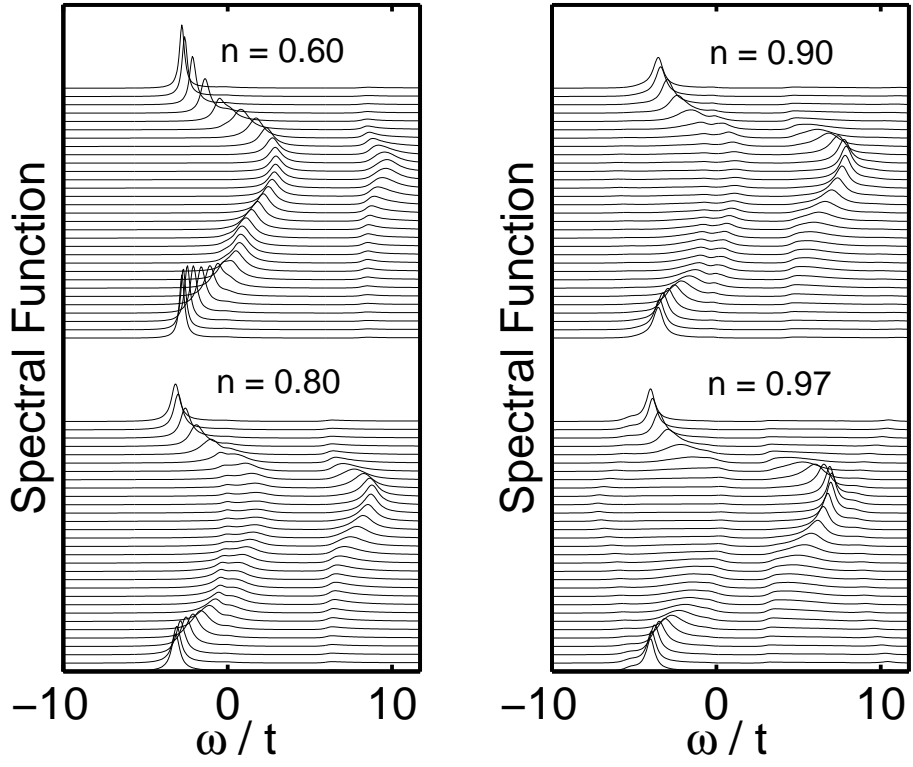}
\caption{Doping dependence of the spectral function in the 2D Hubbard model
computed using the local cluster approach for $T=0.25t$ and fillings
of $n= 0.97$, $n=0.90$, $0.80$, and $n=0.60$.
}
\label{sp25}
\end{figure}

Shown in Fig. (\ref{sp25}) is the electron spectral function at high
temperature, $T=.25t$ for $n=0.97$, $n=0.90$, $n=0.80$, and $n=0.60$.
Several features are evident:  1) the chemical
potential moves further into the LHB as the filling
decreases, 2) no coherent peaks exist near the chemical potential
in the lightly doped regime, $.9<n<1$, 3) in the dense or weakly interacting
regime, sharper features appear, 4) each state in the FBZ has spectral weight
both above and below the chemical potential as dictated by Mottness,
 5) the Mott gap remains intact but moves to higher energy as the doping
increases, and 6) at $(\pi,\pi)$, the UHB carries most of
the spectral weight regardless of the filling.  In the underdoped regime, the characteristic
width of each $\vec k$ state is of order $t$ and even
much larger near $(\pi,0)$.  Such broad spectral features in the underdoped
regime are seen experimentally\cite{hassan} and arise in this context because
$\Im\Sigma(\epsilon_F)\ne 0$ as shown in Fig. (\ref{chempot}). As a consequence, there 
is no sharp criterion for unit occupancy of each state in the FBZ. 
 Because the total spectral weight of each $\vec k$ state is
unity, however, and each state lives both below and above the chemical
potential, the charge carried by the piece of the state lying below the chemical
potential is less than unity.  That is, each electronic state
is fractionalized.  In the heavily overdoped regime, the splitting
of the spectral weight above and below the Mott gap, is highly suppressed.  As
illustrated in Fig. (\ref{sp25}), most of the spectral weight resides in the LHB
for a filling of $n=0.60$.  As a consequence, Mottness vanishes
in the overdoped regime.  In addition, in the heavily overdoped regime,
$\Im\Sigma$ acquires the characteristic $\omega^2$ dependence indicative
of a Fermi liquid.  Hence, with our method we are able to recover
the key characteristics to the transition 
to the traditional Fermi liquid state, namely 1) a vanishing of spectral weight in
the UHB and 2) $\Im\Sigma\approx \omega^2$ near the chemical potential.

\noindent

\begin{figure}
\begin{center}
\includegraphics[height=10cm]{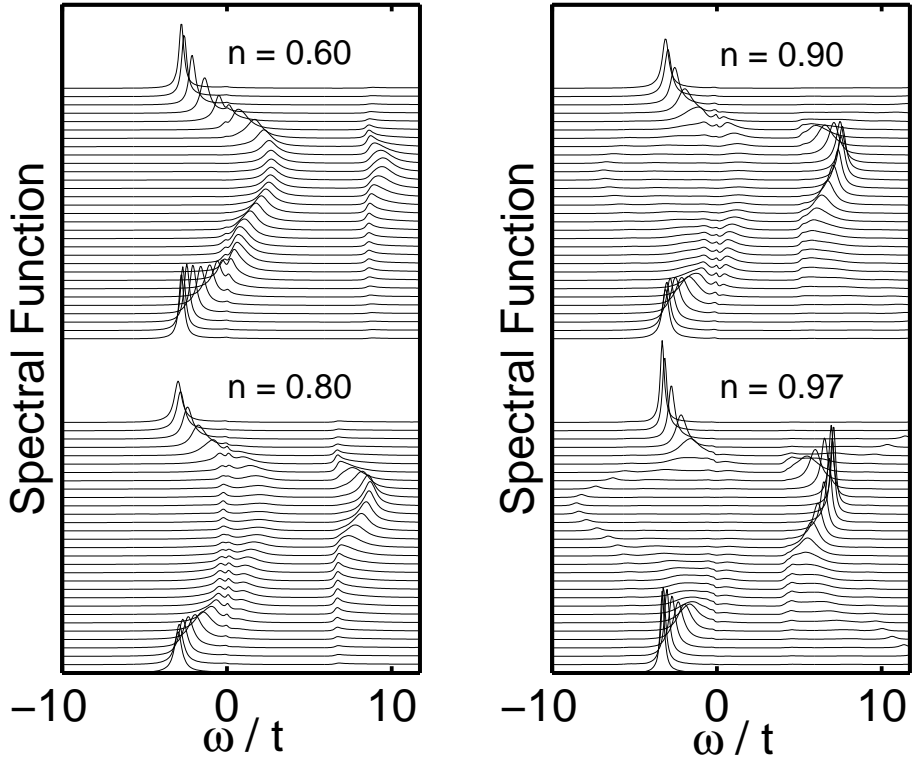}
\caption{Doping dependence of the spectral function in the 2D Hubbard model
computed using the local cluster approach for $T=0.07t$ and fillings
of $n= 0.97$, $n=0.90$, $0.80$, and $n=0.60$.
}
\label{sp07}
\end{center}
\end{figure}

Does new physics emerge at low temperatures?
Fig. (\ref{sp07}) depicts
the spectral function computed at $T=0.07t$.  At this relatively
low temperature, two new features emerge.  First, in the underdoped
regime, the spectral weight appears to be suppressed at the chemical
potential. Whether this gives rise to a
pseudogap will be investigated in the next section.
Second, at $n\approx 0.8$ especially in the vicinity
of the $(\pi, 0)$ point, the band becomes 
almost dispersionless and seems to split into two sub-bands.  
Instead of a strong coherence peak, at the chemical potential
we observe a weak maximum adjacent to a region
with depleted spectral weight which forms
a 'channel' immediately above the chemical potential.  To understand the 
importance of the low-temperature features, it is expedient to compute 
the density of states.
\noindent

\subsubsection{Pseudogap without preformed pairs or global symmetry breaking}

To investigate
the possibility of a pseudogap\cite{pgg1} in the lightly doped regime, we integrate the spectral function over momentum
to obtain the single particle density of states (DOS) at high and 
low temperatures.  
\begin{figure}
\begin{center}
\includegraphics[height=12cm]{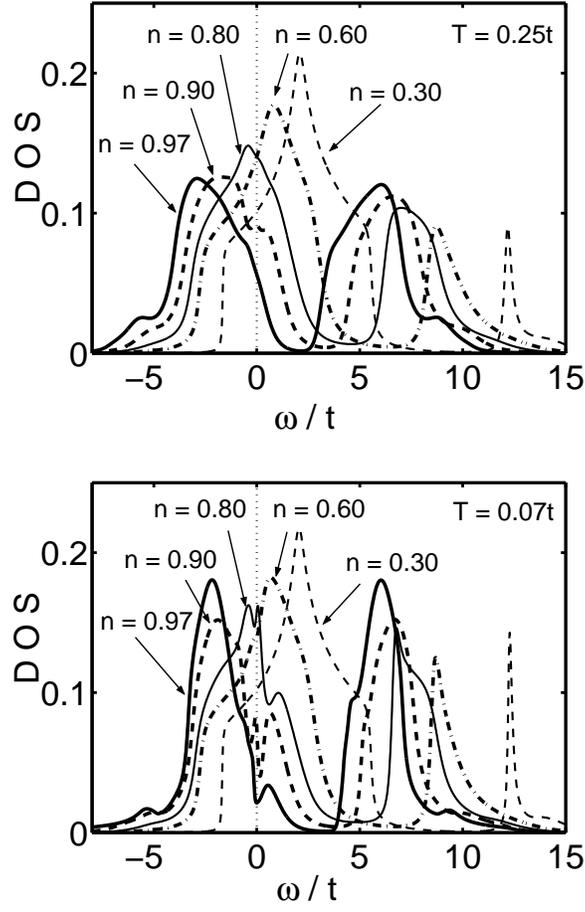}
\caption{Density of single particle states for
$T=0.25t$ and $T=0.07t$, $U=8t$ for the fillings shown.  No pseudogap
exists at high temperature.  At low $T$ and low doping levels, a pseudogap emerges at the chemical potential but  moves above it at an intermediate doping level. In the overdoped regime, the pseudogap vanishes entirely
and a weakly-interacting system is recovered.}
\label{pseudo}
\end{center}
\end{figure} 
\noindent
Displayed in Fig. (\ref{pseudo}) is the DOS for $T=0.25t$ and $T=0.07t$ for several fillings. As is evident, no local minimum of DOS exists at the chemical potential at high temperature, $T=0.25t$.  Features which emerge even at high temperature are the reshuffling of spectral
weight from above the charge gap to below as the filling is changed and also a movement of the Mott gap to higher energies.  Note that even at $n=0.30$ the Mott gap is still present, though almost all of the spectral weight now resides in the LHB
which closely resembles the non-interacting density of states. This is further
evidence that we correctly recover Fermi liquid theory as $n\rightarrow 0$.
What about low temperature?  The lower panel of Fig. (\ref{pseudo}) demonstrates
that a pseudogap forms in the DOS for $\delta\approx 0$.  The vertical
line at $0$ indicates that the pseudogap occurs precisely at the chemical
 potential.  Similar qualitative results based on a cluster
 method have been obtained
by Maier, et. al.\cite{jarrell}, except their pseudogap is slightly displaced
above $E_F$.
In contrast, in the analysis of Haule, et. al.\cite{tj}, the DOS has a negative slope through $E_F$ (as dictated by the proximity to the Mott gap) but never acquires a local minimum at $E_F$ indicative of a
true pseudogap.  Because the pseudogap exists below some characteristic 
temperature and vanishes at higher doping, the result obtained here
 is non-trivial and highly reminiscent of experimental observations
in the cuprates\cite{timusk}.
What is its origin?  Incoherence ($\Im\Sigma\ne 0$ at $E_F$) is central,
 though it is not a sufficient condition\cite{tj}( see Fig. 2) for a pseudogap.
From Fig. (\ref{pseudo}), the pseudogap remains intact up to $\delta=0.20$ but simply moves to higher energies as does the Mott gap.  This is telling because in $d=\infty$,
no\cite{kotliar} pseudogap exists but a Mott gap is present.  Absent from $d=\infty$ but present in any lattice of finite connectivity are true short-range correlations.
We argue then that the pseudogap is the nearest-neighbour analogue of the
on-site generated Mott gap.  The energy scale for nearest-neighbour
interactions scales as $t^2/U$.  Hence, if our hypothesis is 
correct, we expect the pseudogap to diminish as
$U$ increases.  The evolution
shown in Fig. (\ref{ueffect}) indeed demonstrates
that, at finite temperature, the pseudogap does vanish as $U$ increases.  Hence,
we can assert with certainty that
correlations on neighbouring sites do in fact create
a depletion in the density of states.  The energy scale $t^2/U$ is typically associated with antiferromagnetic spin fluctuations.  To explore whether such
processes have the right doping dependence to explain the origin of the pseudogap, we display in Fig. (\ref{jeffx}) the $x-$dependence of $J$ computed from
Eq. (\ref{Jequation}).  As is evident, $J$ is only weakly doping dependent in the underdoped regime and hence lacks the strong doping dependence needed to explain the
pseudogap.  This trend is consistent with that of Jarrell and co-workers\cite{jarrell} who have observed that the pseudogap persists even if antiferromagnetism is killed.  Fig. (\ref{jeffx}) also indicates that $J_{\rm eff}$ computed
as the energy difference between the nearest-neighbour singlet and triplet states vanishes at $x=.8$.  This is not an accident.  Nearest-neighbour spin fluctuaions should desist when no nearest-neighbour sites remain singly occupied. On average, this obtains at a filling of $1/5$ or $x=.8$, precisely the doping value
found here.  Hence, it is not a surprise that the doping dependence of $J$ is weak in the underdoped regime.  
 \begin{figure}
\begin{center}
\includegraphics[height=6cm]{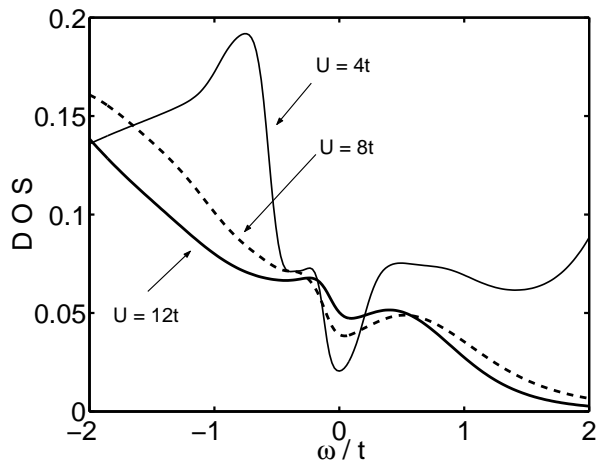}
\caption{Density of single particle states for
$T=0.1t$, for $n=0.95$ and three values
of the on-site interaction:  $U=12t$, $U=8t$, and $U=4t$.  The gradual
vanishing of the pseudogap as $U$ increases offers direct confirmation
that the energy scale for the pseudogap is set by $t^2/U$. }
\label{ueffect}
\end{center}
\end{figure}

\begin{figure}
\begin{center}
\includegraphics[height=6cm]{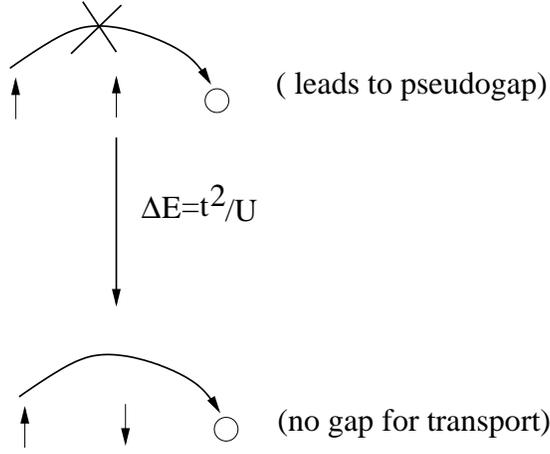}
\caption{Local three-site configurations in which spin-blocking leads to a pseudogap. In the upper state, transport directly between the up spin two sites away from the hole and the hole is not possible. Only hole transport via a two-step process is possible. As the amplitude for hole transport is a superposition of all such processes, a pseudogap develops. The only way to overcome the spin-blockade is for the spin neighbouring the hole to flip.  This process costs an energy $t^2/U$.  Once the spin is flipped, there is no barrier for transport.}
\label{pseudo1}
\end{center}
\end{figure}

\begin{figure}
\begin{center}
\includegraphics[height=6cm]{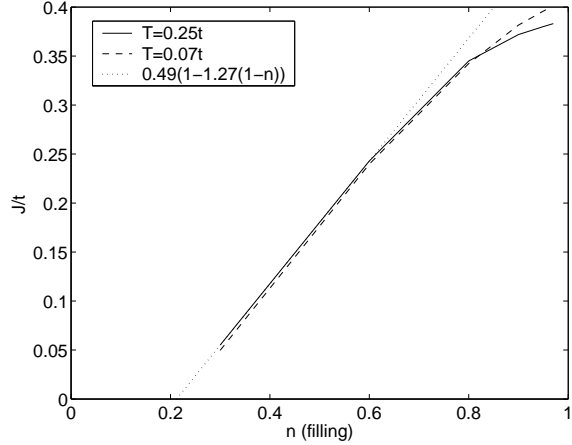}
\caption{Effective nearest-neighbour $J$ as a function of filling.   $J$ vanishes when four out of five sites is empty.  Beyond this concentration,
nearest-neighbour spin fluctuations are not possible.}
\label{jeffx}
\end{center}
\end{figure}

What then is the cause of the physics of the pseudogap? Any two-step process involving the UHB scales as $t^2/U$.  Consider the explicit three-site terms that appear away from half-filling
\beq
H_{\rm 3B}=-\frac{J}{4}\sum_{i\delta\ne\delta',\sigma}\left(
\tilde{c}^\dagger_{i+\delta,\sigma}\tilde{\eta}_{i,i+\delta'\sigma,-\sigma}
-\tilde{c}^\dagger_{i+\delta,-\sigma} \tilde{\pi}_{i,i+\delta',\sigma}\right)
\label{3B}
\eeq
in an expansion in $t/U$, where the tilde represents full projection
onto the LHB.  It is precisely these terms that thwart the equivalence between the so-called $t-J$ and Hubbard models in the large $U$ limit away from half-filling\cite{eskes}.  
 This term represents the motion (strictly in the LHB) of a hole in
a spin background.  We argue that such terms are involved in the pseudogap.
The mechanism is as follows.  Consider placing a single
hole in a Mott insulator. Unlike a site neighbouring the hole, a singly-occupied site two lattice sites away must temporarily doubly occupy one of its neighbours if it is to move to the hole.  For this to be possible, the electrons on neighbouring sites must have opposite spins.   The matrix element for such a two-step
process is $t^2/U$ and described by the three-site terms written above. For sites with the incorrect spin alignment, a local spin fluctuation must obtain for
the three-site hopping to occur.  The energy barrier for this process is
$t^2/U$.  It is from those local three-site configurations in which the spins are incorrectly aligned that the pseudogap arises as illustrated in Fig. (\ref{pseudo1}).  Simply invoking spin fluctuations is insufficient to explain the origin of the pseudogap as spin fluctuations alone cannot give rise to transport.  However, spin fluctuations
can make it impossible for an electron two sites away from a hole to transport.
Hence, spin fluctuations in the context of three-site hopping can overcome the
local spin blockade (or spin gap) that exists in doped Mott insulators.
As this effect is entirely local, the pseudogap is the nearest-neighbour
analogue of the Mott gap: neighbouring sites with a parallel arrangement of the spins experience an energy barrier equal to $t^2/U$ for charge transport.  Can the doping value at which this process vanishes
be estimated?  On this account, the pseudogap should be related to the joint probability that a neighbouring three-site configuration consiting of a hole and two sites with spin parallel electrons exists.  The minimum constraint however is simply that each site has on average one hole as its immediate neighbour,
roughly $x=0.25$ for a square lattice.
Hence, the pseudogap is of the form $t^2/UP(x)$, where $P(x)$ 
determines the probability that hole transport involves double occupancy and 
consequently, is a steadily decreasing function of $x$ vanishing at $x_{\rm crit}$.  From the estimate given above, it is likely that $x_{\rm crit}$ is closer
 to the end of the superconducting dome than it is to optimal doping. 
 
It is common\cite{eskes,loop} to approximate the three-site terms considered here as $xKJ$, where $K$ is treated as
the lattice connectivity.
Consequently, the effective nearest-neighbour exchange
interaction is doping dependent and given by $J_{\rm eff}=J(1-xK)$ which
vanishes at $x_{\rm crit}=1/K\approx .25$\cite{loop} for a square lattice. However, as Fig. (\ref{jeffx}) demonstrates, $J_{\rm eff}$ vanishes
at $x=.8$ not $x=.25$.  It is likely that the discrepancy found here arises from
the fact that the 3-body terms in Eq. (\ref{3B}) cannot, in any real sense,
be rewritten as an effective spin-exchange.  In fact,  at $x=.8$, every site has on average four neighbouring
holes.  At this concentration, nearest-neighbour spin fluctuations are not possible; hence, $J_{\vec eff}$ should vanish.  

\subsubsection{Heat Capacity}

\begin{figure}
\begin{center}
\includegraphics[height=6cm]{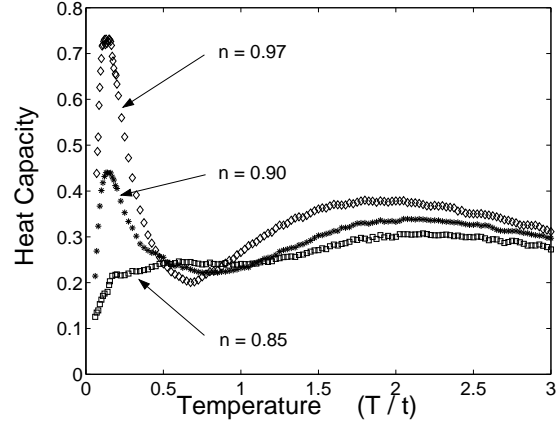}
\caption{Heat capacity computed for three fillings computed by numerically 
differentiating the internal energy obtained from the average of Eqs. (\ref{en1}) and (\ref{en2}). For all three fillings shown, $U=8t$. }
\label{hcap}
\end{center}
\end{figure}

Evidence for the pseudogap is also found from the heat capacity\cite{loram}.
We computed the heat capacity numerically from the internal energy
\beq
C(T) = \frac{1}{N} \frac{dE}{dT}.
\eeq
The energy per site is the sum of the kinetic term and the interaction
term
\beq
\frac{E}{N} = -2t\langle c_i^{\dagger}c_i{\alpha}\rangle
             + U \langle n_{i\uparrow} n_{i\downarrow} \rangle,
\eeq
where $t = 2dt_0$ ($t = 4t_0$ in two dimensions).
The double occupancy can be expressed as
\beq
\langle n_{i\sigma} n_{i\bar{\sigma}} \rangle =
\langle c_{i\sigma}^{\dagger} \eta_{i\sigma}\rangle =
\frac{n}{2} - \langle \eta_{i\sigma} c_{i\sigma}^{\dagger}\rangle.
\eeq
The correlations can be expressed in terms of the Green function as
\beq\label{en1}
\frac{E}{N}& =& \frac{n}{2}U +
\int\frac{d\omega}{2\pi}\int\frac{d^2k}{(2\pi)^2}
(1-f(\omega))[2t\alpha(k)\frac{-1}{\pi}Im(S_{11}+2S_{12}+S_{22}) -
U\frac{-1}{\pi}Im(S_{12}+S_{22})].
\eeq
Equivalently, we can use the equation of motion for the Green function
$\langle\langle \eta_i(\tau), c_j^{\dagger}(\tau') \rangle\rangle$ and we
obtain an alternate expression
\beq\label{en2}
\frac{E}{N} = \frac{n}{2}U -(1-\frac{n}{2})\mu -
\int\frac{d\omega}{2\pi}\int\frac{d^2k}{(2\pi)^2}
(1-f(\omega))(\omega - t\alpha(k))\frac{-1}{\pi}Im(S_{11}+2S_{12}+S_{22})
\eeq
for the energy per particle.  We found that the difference between Eqs. (\ref{en1}) and (\ref{en2}) wqs within our numerical errors.  Consequently, in our
final calculations of the heat capacity shown in Fig. (\ref{hcap}), we averagedthe two results.  In the 1D and 2D Hubbard models at half-filling\cite{stanescu1}, two peaks exist in the heat capacity.  The high temperature peak corresponds to charge excitations and the low-tempeature peak to spin physics.  As is evident, the same separation of energy scales persists even in the doped case.  However, for $n<0.9$, we find that the spin peak vanishes and merges by the charge excitation spectrum.  This dramatic change represents a possible termination of Mott-dominated physics and the onset of more Fermi liquid behavior.  Bonca and Prelovsek\cite{bonca} observed the identical trend in their exact diagonalisation study of
of a $4\times 4$ system.  This agreement lends further credence to our method.
Another trend evident from Fig. (\ref{hcap}) is that the extrapolated $T=0$ value of the heat capacity in the underdoped regime, $n>0.9$ is lower than that at
$n=0.85$.  This represents a loss of spectral weight at low energies as would be the case once a pseudogap opens.  Hence, the thermodynamics also corroborate the existence of a pseudogap.

\noindent

\subsubsection{Possible time-reversal symmetry-breaking}

\begin{figure}
\begin{center}
\includegraphics[height=6cm]{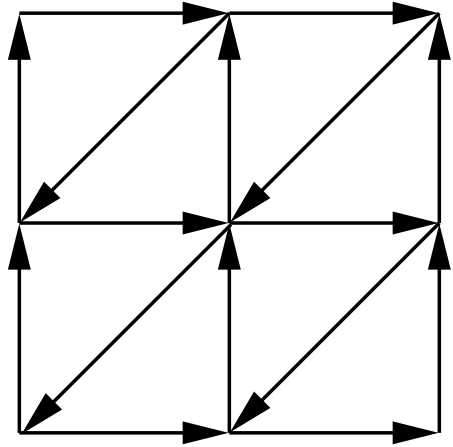}
\caption{Current pattern for the motion a hole in a doped Mott insulator
that preserves t-reversal symmetry along $x=\pm y$ but violates it along
the canonical $x$ and $y$ axes. Each lattice represents a copper site.
Hence, translational symmetry is preserved. The diagonal current is chosen
such that the net current in each plaquette vanishes.}
\label{current}
\end{center}
\end{figure}

We have shown that in general, a pseudogap exists in a doped Mott insulator
without invoking symmetry breaking of any kind.  Nonetheless, we entertain
the possibility that in analogy with antiferromagnetism and the Mott gap,
perhaps some broken symmetry state obtains at lower temperature as a result 
of the pseudogap found here.  Our argument should be construed as a conjecture
and hence is entirely speculative.  It by no means underlies the calculations we presented here.
 Loram and colleagues\cite{loram} have 
argued that for $x<x_{\rm crit}$, a  glassy phase with
an Edwards-Anderson order parameter obtains.
While the experimental evidence for a glassy phase extending to $x_{\rm crit}$
is not clear, recent circular dichroism experiments\cite{campu} point to time-reversal as the relevant symmetry
that is broken in the pseudogap phase.  However, this symmetry is broken only
along the $(\pi,0)$ and $(0,\pi)$ directions and not along $(\pi,\pi)$.
Should these results endure, they will provide a benchmark for measuring
the validity of the numerous proposals for the pseudogap\cite{pg1,pg11,pg2,pg3,pg4}.
In the context of the view put forth here, we must determine how
purely nearest-neighbour correlations can give rise to a breaking of time-reversal symmetry only along the canonical $x$ and $y$ axes but not along $x=y$.
Consider the three-body term in Eq. (\ref{3B}).  This term generates
correlated motion of a hole among  nearest-neighbour sites, that is,
local currents.  In analogy
with the local moments that order antiferromagnetically $T=0$ as a result of the Mott gap,
we propose that the currents may order 
in the pseudogap phase below $T^\ast$.  Experimentally\cite{campu}, translational symmetry is preserved
in the pseudogap phase.  Hence, 
staggered orbital currents are not possible as they automaticaly
result in a doubling of the unit cell within a single-band model\cite{campu}.  Further, experimentally\cite{o1,o3,newshen}, there is no evidence that physics beyond a single-band is relevant 
to the cuprates.  Consequently, any current pattern must preserve translational symmetry within a single band model.  Only one option remains: the currents order below
some characteristic temperature, $T^\ast$, along the canonical x and y axes.
To ensure that the net current along $x=y$ vanishes, a compensating diagonal current must be present as depicted in Fig. (\ref{current}).  This current pattern
can be obtained from the most recently proposed pattern of Simon and Varma\cite{varma2} by simply integrating out the oxygen sites.  Hence, despite 
claims to the contrary\cite{varma2}, it is entirely possible to generate
a translationally invariant current pattern within a 1-band model that is consistent with the
experimental observations.  In the corrected pattern of Varma and Simon\cite{varma2}, the oxygen sites do nothing except produce a diagonal current which ensures that the total current in each plaquette vanishes. That the oxygen sites can
be integrated out is certainly consistent with the now well accepted 
work of Zhang and Rice\cite{zr}.  Nonetheless, our work does not hinge on the current pattern shown in Fig. (\ref{current}) being the origin of the pseudogap.
However, insofar as such a pattern obtains
entirely from local nearest-neighbour physics,
 it is consistent with our finding that the pseudogap
is the second harmonic of the Mott gap. Should further
experiments confirm the presence of t-violation, then a more microscopic
investigation of the origin of the current pattern shown in Fig. (\ref{current})
will be warranted.   Of course, a current pattern of the type proposed here can
only be obtained (if at all) from a Hubbard model if next-nearest neighbour hopping 
is included.

\noindent

\subsubsection{Spectral Weight Transfer}

\begin{figure}
\begin{center}
\includegraphics[height=6cm]{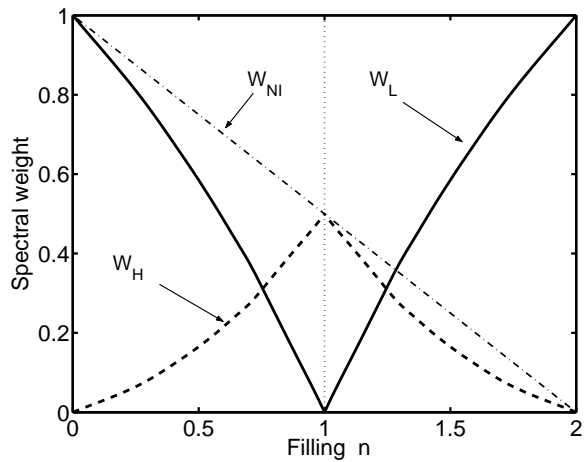}
\caption{High ($W_H$) and low ($W_L$) spectral weight as a function of filling. $W_{\rm NI}$ is the spectral
weight in the non-interacting system.}
\label{sweight} 
\end{center}
\end{figure}

To quantify the spectral weight transfer evident in Fig. (\ref{pseudo}),
we compute
the high and low spectral weight by integrating the DOS from a value of the  energy inside the Mott 'gap' which
minimizes the DOS to $\infty$ ($-\infty$ for electron doping) and from $\mu$ to that fixed energy, respectively.   The results shown in Fig. (\ref{sweight})
demonstrate that the initial spectral weight in the UHB which is $1/2$ at $n=1$
all moves to low energies as the filling decreases, as in observed experimentally\cite{cooper,uchida1}. 
The integrated spectral weight has been normalized per spin.   The same 
is true for electron doping ($n>1$).  Further, the curvature
of the low energy spectral weight is positive as a function of doping
in agreement with earlier results\cite{eskes} on the 1D Hubbard model.  This signifies
that the integrated low-energy spectral weight increases faster than $2x$.  The additional
low-energy spectral weight above that dictated by state 
counting (see Fig. (\ref{fig1}) arises from
virtual excitations between the lower and upper Hubbard bands. Such virtual transitions arise from the three-site terms discussed previously..  That the number of low energy
degrees of freedom arise from such high energy processes further attests to the inseparability of 
the low and high energy degrees of freedom in a strongly correlated system. For contrast, the spectral
weight transfer for a non-interacting system ($W_{NI}$) is shown as well. This behavior is expected for a doped band insulator.  That Mottness leads to such
a drastic deviation from the non-interacting result is a direct consequence
of state fractionalization.  That is, each state has spectral weight both above and below
the chemical potential (see Fig. (\ref{sweight}).

\subsubsection{Hall Coefficient}

Experimentally,
the Hall coefficient in the cuprates is in general positive in the lightly
hole-doped regime, scales as $1/x$ in the vicinity of half-filling but
falls off faster than $1/x$ for $x\approx .1$ and in some instances changes 
sign\cite{ongreview,tagaki,uchida} typically around $x=.25$. Although the Hall coefficient is in general
temperature dependent as emphasized extensively by
Anderson\cite{hallanderson}, the Takagi, et. al. experiments\cite{tagaki}
indicate that the zero-crossing doping level in La$_{2-x}$Sr$_x$CuO$_4$ (LSCO)
is only weakly temperature dependent. Consequently, we will focus solely
on the doping dependence of the Hall coefficient since the existence of 
the zero-crossing is only weakly doping dependent.
Nonetheless, because the 
sign change is not universally observed in all the cuprates, the general
conditions under which a sign change of the Hall coefficient should be observed in a doped Mott
insulator
have not been formulated.  In addition, there have been numerous theoretical treatments of the Hall coefficient. For example, perturbative schemes\cite{ongreview,levin,trugman,doniach}
lead to a sign change of $R_H$ and hence offer a possible
explanation for the deviation from $1/x$.  However,
because perturbation theory is constrained by Luttinger's theorem\cite{lutt}
to yield a Fermi surface occupying half the first Brillouin zone (FBZ) at half-filling, such approaches fail to recover the experimentally observed\cite{ongreview}
divergence of $R_H$ at half-filling.  In strong-coupling
calculations, some have obtained a sign change\cite{castillo,rojo,zotos}
while others\cite{imada} predict that $R_H<0$ for all hole dopings. 
In addition, others\cite{shastry,schm} have reached the counterintuitive 
conclusion that $R_H$ does
change sign, but the Fermi surface is
closed for all $x>0$. In such studies, it was assumed
that the doped Mott insulator is described by doping the diamond-shaped
Fermi surface of the weakly-interacting system, an assumption
clearly not borne out by experiment\cite{lsco3,o1,o3}. 

On simple grounds, however, the general doping dependence of the Hall coefficient can be easily
deduced.  Consider a lightly-doped Mott insulator in which the Hall coefficient is initially
positive.  In the heavily overdoped regime where the system is weakly interacting, the Fermi
surface must be closed and hence electron-like; thus $R_H<0$. We can deduce the doping level
at which the transition from an open to a closed Fermi surface occurs by appealing to the spectral
function of a Mott insulator.  As Figs. (\ref{fig4.1a}) and 
(\ref{nk}) illustrate, at half-filling
every $\vec k-$ state in the FBZ has some spectral weight.  Because the chemical potential (see Fig. (\ref{chempot})) simply
moves down through the LHB upon doping, hole-doping simply depletes the spectral weight
in the LHB.  When half the spectral weight in the LHB is removed, the surface separating
occupied from empty states must have zero curvature.  As this surface defines the Fermi surface,
then its curvature should be related to the sign of the Hall coefficient. Consequently, the critical doping level at which $R_H=0$ is 
determined by the doping level at which half the spectral weight in the LHB is depleted.  Consider
the $U=\infty$ limit in which the spectral function is momentum independent.  Each state in
the LHB carries the weight $1-n/2$.  Consequently, the fraction
of the spectral weight depleted upon hole
doping is $n/2/(1-n/2)$. When this quantity equals $1/2$, $R_H$ should
vanish.  The solution
to 
\beq
\frac{n/2}{1-n/2}=\frac{1}{2}\quad\quad {\mathrm\, lower\,
 limit\, on\,
\,the\, filling\, at\, which\,} R_H=0
\eeq
 is $n=2/3$ which
is a strict lower bound for $n_c$ and
is precisely what simulations as well
as complicated series expansions in the infinite-$U$ limit\cite{zotos,shastry} obtain.  Consequently,
if a Mott insulator possess a sign change in the Hall coefficient, it must occur for $x<.333$.

\begin{figure}
\begin{center}
\includegraphics[height=8cm]{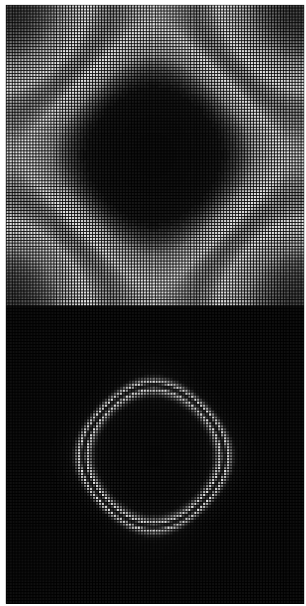}
\caption{(Color) Spectral function  in the first Brillouin
zone evaluated at the chemical potential for 
a filling of $n=0.97$ (top panel) and $n=0.30$ (bottom panel) with $U=8t$
and $T=.25t$. Color scheme: blue= minimum (zero) spectral weight,
 red= maximum, yellow= intermediate values.  In the underdoped regime, the spectral function has broad features at the Fermi level.
A sharp Fermi surface emerges in the overdoped regime.}
\label{fs25}
\end{center}
\end{figure}

Shown in Fig. (\ref{fs25}) is the spectral
function in the FBZ evaluated at the chemical potential for  
$U=8t$ and $T=.25t$. The upper panel corresponds to $n=0.97$ and the lower to $n=.3$.  As is clear, in the lightly-doped regime, the Fermi surface
is hole-like and the spectral features are broad indicating an absence of well-defined quasi-particles indicative
of an incoherent metal as is seen experimentally\cite{o1,o3}.  The source
of the incoherence stems from the self-energy shown in Fig. (\ref{chempot}) which remains
constant at the Fermi level at $n=0.97$.  This leads necessarily to 
a violation of Luttinger's theorem. In fact, the Fermi surface  (defined by the 
maximum in the spectral function) volume
at $n=0.97$ is roughly 30$\%$ larger than the Luttinger volume.  In the overdoped regime, the self-energy has the characteristic
$\omega^2$ dependence of a Fermi liquid and hence we recover Luttinger's theorem as the sharp
spectral features in Fig. (\ref{fs25}) reveal for $n=0.30$.   Our results
indicate a smooth crossover between the lightly doped regime and overdoped regimes where Luttinger's
theorem is reinstated.
 
\begin{figure}
\begin{center}
\includegraphics[height=8cm]{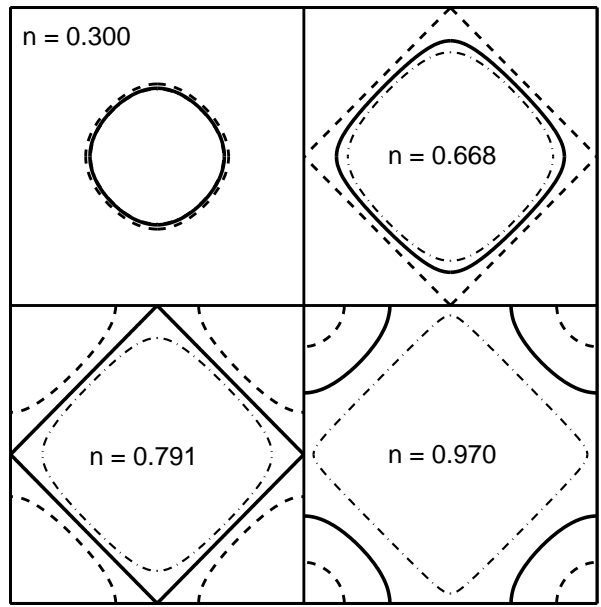}
\caption{Fermi surface in the static approximation for four fillings as indicated and three different values of $U$: 1) solid line, $U=8t$, 2) dashed line,
$U=1000t$ and 3) dashed-dotted line, $U=0$, the non-interacting limit.}
\label{static}
\end{center}
\end{figure}

However, broad spectral features are not the only contributor to the violation
of Luttinger's theorem.  Consider the static approximation in which the self
energy in Eq. (\ref{greenEq}) is explicitly set to zero. The details of this 
level of theory are derived in Appendix E. At this level of theory, the spectral function for the LHB and UHBs correspond
to a series of $\delta-$ functions.
Nonetheless, the bands generated do not describe Fermi liquid quasi-particles
because each $\vec k-$ state still has spectral
weight both below and above the Fermi level.  Relative to the dynamical
results, we find that the topology and volume of the 
Fermi surface do not change as revealed by Fig. (\ref{static}).  The solid line corresponds to 
$U=8t$, dashed line to $U=1000t$ and dashed-dotted to $U=0$. Clearly shown in Fig. (\ref{static}) is the evolution
from a hole to an electron-like Fermi surface at critical doping levels of $.791$ for $U=8t$ 
and $.668$ for $U=1000t$. 
The critical concentrations at which the curvature
of the Fermi surface changes sign corresponds 
to $x_c=0.668$ and $0.791$ for $U=1000t$ and $U=8t$, respectively. As anticipated,
these values are less than $x_c=2/3$. While for $U=8t$, $x_c$ is remarkably close to the 
$x_{\rm crit}=.19$ of Loram\cite{loram}, it is unclear whether the 
closing of the pseudogap is always accompanied with a sign change of the Hall
coefficient. However, a sudden sign change would certainly explain
the appearance of the peak in the density of states shown in 
Fig. (\ref{pseudo})once the pseudogap vanishes.  Three additional
features are apparent. First,
at small concentrations, regardless of $U$,
all Fermi surfaces are electron-like and coincide with the non-interacting
limit.  An analytical proof of this result in given in Appendix E. Second, at intermediate fillings, the Fermi surface
in the interacting system is hole-like as opposed to electron-like
in the non-interacting system.  Finally, the area of the FS for $U=8t$ 
and $n=1-x=.97$ is clearly larger than that dictated by
 Luttinger's theorem,
$2\pi^2(1-x)=1.94\pi^2$.  From the maximum in the spectral function, we find that
the experimental value for the FS area in LSCO\cite{newshen} for $n=.97$ is $2.06\pi^2$ which represents a non-trivial
8$\%$ deviation from the Luttinger result. Such a large deviation cannot be attributed to experimental uncertainty (at most 2-3$\%$)\cite{uncer}. Hence, while the FS computed here is clearly larger ($2.5\pi^2$) than the experimental value, both
 are in qualitative agreement that Luttinger's theorem is violated in the underdoped regime.
Of course, improved quantitative agreement with experiment can be obtained
by including band parameters such as a next-nearest neighbour hopping interaction, $t'$ and a doping dependent $U$.  Hence, we see that even though the UHB and LHBs are sharp, the Luttinger volume
is not preserved.  In Appendix E, we show explicitly that the source of this breakdown stems from
the bifurcation of the spectral weight of each $\vec k-$state into a high and low energy part.
Consequently, removing 
a single electron is no longer accomplished simply by removing a single $\vec k-$state as 
a result of the breakdown of the band insulator sum rule. 
 Hence, a key consequence of
Mottness in 2D is a violation of Luttinger's theorem for $n\ne 1$
as additional extensive numerical work attests\cite{puttika,ogata,jarrell2,qmc2,amv}.  In the heavily overdoped regime (see Fig. (\ref{static}), the spectral lies predominantly
in the LHB and hence one hole=one $\vec k-$ state and Luttinger's theorem is reinstated.
 Consequently, under hole doping,
the hole and electron regimes are fundamentally asymmetrical
as emphasized by Hirsch\cite{hirsch}.

\begin{figure}
\begin{center}
\includegraphics[height=6cm]{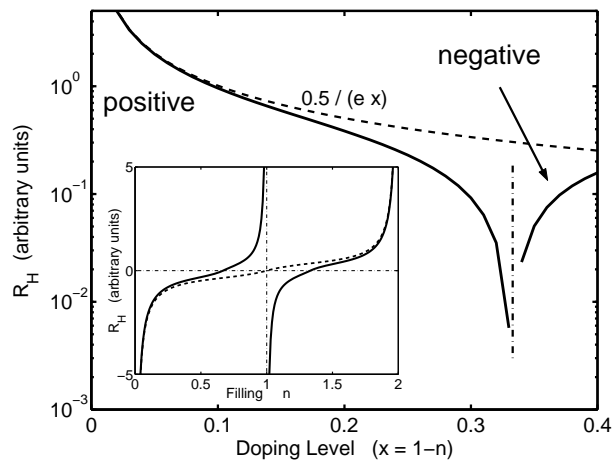}
\caption{Hall coefficient as a function of doping
using the static approximation for the Hubbard
operators with $U=1000t$.  The inset shows that there is an antisymmetry
between electron and hole doping. In the inset, the solid line
corresponds to $U=1000t$ and the dashed line to $U=0$.  Both show clearly that the deviation
from $1/x$ is induced by the sign change rather than a liberation
of charge.} 
\label{hall}
\end{center}
\end{figure}

To compute the Hall coefficient,
\begin{equation}
R_H = \sigma_{xyz}/\sigma_{xx}^2
\label{rh}
\end{equation}
we work within Boltzmann transport theory in which\cite{PA1}
\begin{equation}
\sigma_{xyz} = \frac{e^3\tau^2}{\hbar\Omega c}\sum_{\vec{k}}
v_x (\vec{v} \times \vec{\nabla}_{k})_z v_y
(-\frac{\partial f}{\partial \epsilon_k})
\label{sigmaxyz}
\end{equation}
\begin{equation}
\sigma_{xx} = \frac{e^2\tau}{\Omega}\sum_{\vec{k}}v_x^2
(-\frac{\partial f}{\partial \epsilon_k})
\label{sigmaxx}
\end{equation}
Here, 1/$\tau$ is the scattering rate,
$\Omega$ the
volume, and $f$ the Fermi distribution function. Our use of the Boltzmann
equation should suffice as long as the interactions dominate which
is certainly true in the case of interest, $T\ll U$.  In using the Boltzmann
approach, it is easiest to work in the large $U$ limit because in this case,
the spectral function is independent of momentum.  
Consequently, we used the static energy bands for $U=1000t$ and compute
$R_H$ using Eqs. (\ref{rh})-(\ref{sigmaxx}).
Fig. (\ref{hall}) demonstrates that the sign of the Hall
constant is consistent with the curvature of the Fermi surfaces
shown in Fig. (\ref{static}).  In addition, the deviation from $1/x$ in the region close
to $x_c=1-.668$ is tied to the impending sign change.  
As the inset illustrates, $R_H$ diverges at half-filling and
changes sign for both electron and hole
doping in contrast to weakly-interacting
scenarios which can yield at most one sign change (dashed line) and no
divergence at $n=1$. Of course, the static approximation does
not include the pseudogap found earlier.
Note that regardless of which model is used for the 
pseudogap, the $(\pi,0)$ regions of the Fermi
surface become gapped. Unless the curvature of the Fermi surface is
modified by the removal of the $(\pi,0)$ regions,
the pseudogap cannot change the sign of the Hall coefficient nor eliminate
the divergence at half-filling. In fact, at the doping level ($x\approx x_c$)
 at which
the removal of the $(\pi,0)$ regions is most likely to affect the 
curvature of the FS, the pseudogap vanishes.  

\section{Final Remarks}

\begin{figure}
\begin{center}
\includegraphics[height=10cm]{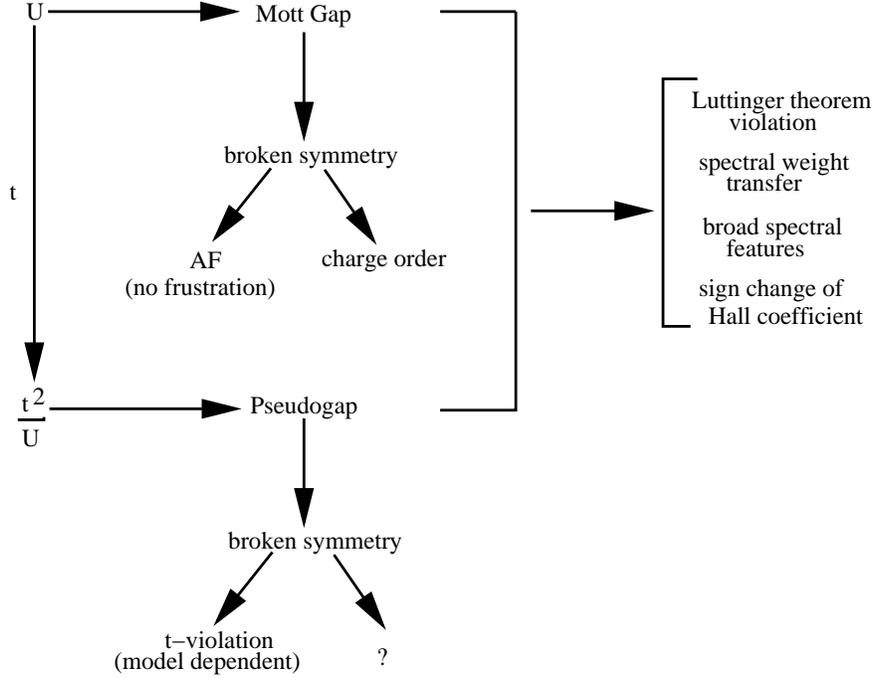}
\caption{Heirarchy of energy scales and the corresponding physical
processes that obtain in a doped Mott insulator.  AF represents
antiferromagentic order.} 
\label{full}
\end{center}
\end{figure}

We have explored here a dynamical method which incorporates the local physics of a doped Mott 
insulator. The summary of our findings is catalogued in Fig. (\ref{full})  Physics on the Mott gap scale $U$ as well as the nearest-neighbour interaction scale, $J$,
play several key roles.  The Mott scale, $U$, sets the energy range for spectral weight transfer
and leads to a breakdown of the band insulator sum rule.  This ultimately leads
to a Fermi surface volume that exceeds that dictated by Luttinger's theorem in the underdoped regime.
Reinstatement of the Luttinger volume in the overdoped regime points to a fundamental
asymmetry between holes and electrons in a hole-doped Mott insulator.  An additional role played by the Mott scale is the generation of 
a hierarchy of interactions of increasing range.  The most important of these 
is the 
nearest-neighbour interaction, $J\approx t^2/U$. Spectral weight
transfer across the Mott gap points to an inseparability of high
and low energy scales.  Hence, it is unclear in what sense a
true low-energy theory can be formulated for a doped Mott insulator. We propose that the antiferromagnet that forms in a Mott insulator is distinct from a spin-density wave antiferromagnet.  
Finally, we have found that the $J$ is also responsible for 
the pseudogap.  The pseudogap simply reflects the restricted phase space that 
strongly correlated excitations on neighbouring sites encounter.  
The current
pattern shown in Fig. (\ref{current}) arises from such neighbouring correlations
and could explain the origin of the direction-dependent t-reversal symmetry
breaking observed
in the normal states of the cuprates. 
Three-site correlations, which
lead to a doping dependent spin exchange interaction,
are crucial to the vanishing of the pseudogap
at $x_{\rm crit}$.   Because this state of affairs obtains beyond $x_{\rm opt}$, our proposal
resonantes with that of Loram and colleagues\cite{loram}. 
Finally, our work suggests that doping a Mott insulator gives rise to a heirarchy of 
energy scales all derived from the Mott gap, $U$.  Hence, 
the Mott state
found here has the high energy scale needed to explain the spectral
weight transfer from $2eV$ to the Fermi energy when superconductivity obtains\cite{rubhaussen,marel,bontemps}.  Whether the emergence of successively
lower energy scales as a function of doping can be formulated within
a renormalization group scheme remains an open question in strongly
correlated electron physics.  Nonetheless, it is along these lines that
our current work is directed.

\section{Appendix A: Equations of motion for the two-level operators}

Consider the orthogonalized basis for the two-site problem, $\Phi_n$, Eqs. (\ref{phibasis1},\ref{phibasis2}). The equations of motion
\begin{equation}
i\frac{\partial}{\partial t} \Phi_n = [\Phi_n, H],
\end{equation}
where H is the Hubbard Hamiltonian,  can be easily obtained using the equations of motion for the single-site level operators, Eq. (\ref{singeq}). 
Assuming that $\Phi_n$ is defined on $x$ and $x'$, a superscript $\bar{\alpha}$ will indicate a sum over all the nearest-neighbors of $x$ with the exception of $x'$ divided by the total number of nearest-neighbors. Similarly, $\bar{\alpha}'$ will include all the nearest-neighbors of $x'$ with the exception of x.
Explicitly, we have:
\begin{eqnarray}
i\frac{\partial}{\partial t} FB_S^{\sigma} &=& \left(2\varepsilon_0 - \mu - \frac{\tilde{t}}{2d}\right) FB_S^{\sigma} - \frac{\tilde{t}}{\sqrt{2}}(c_{\sigma}^{\bar{\alpha}} + c_{\sigma}^{\bar{\alpha}'}) BB   
- \frac{\tilde{t}}{\sqrt{2}}(c_{\sigma}^{\dagger\bar{\alpha}} - c_{\sigma}^{\dagger\bar{\alpha}'}) FF^{\sigma} 
\nonumber \\
 &-& \frac{\tilde{t}}{2}(c_{-\sigma}^{\dagger\bar{\alpha}} - c_{-\sigma}^{\dagger\bar{\alpha}'}) FF_S + \frac{\tilde{t}}{2}(c_{-\sigma}^{\dagger\bar{\alpha}} + c_{-\sigma}^{\dagger\bar{\alpha}'}) FF_A    \nonumber \\
&-& \frac{\sigma\tilde{t}}{2}(c_{-\sigma}^{\dagger\bar{\alpha}} + c_{-\sigma}^{\dagger\bar{\alpha}'}) DB_S - \frac{\sigma\tilde{t}}{2}(c_{-\sigma}^{\dagger\bar{\alpha}} - c_{-\sigma}^{\dagger\bar{\alpha}'}) DB_A  ,
\end{eqnarray}

\begin{eqnarray}
i\frac{\partial}{\partial t} FB_A^{\sigma} &=& \left(2\varepsilon_0 - \mu + \frac{\tilde{t}}{2d}\right) FB_A^{\sigma} - \frac{\tilde{t}}{\sqrt{2}}(c_{\sigma}^{\bar{\alpha}} - c_{\sigma}^{\bar{\alpha}'}) BB   
+ \frac{\tilde{t}}{\sqrt{2}}(c_{\sigma}^{\dagger\bar{\alpha}} - c_{\sigma}^{\dagger\bar{\alpha}'}) FF^{\sigma} 
\nonumber \\
 &+& \frac{\tilde{t}}{2}(c_{-\sigma}^{\dagger\bar{\alpha}} + c_{-\sigma}^{\dagger\bar{\alpha}'}) FF_S - \frac{\tilde{t}}{2}(c_{-\sigma}^{\dagger\bar{\alpha}} - c_{-\sigma}^{\dagger\bar{\alpha}'}) FF_A    \nonumber \\
&-& \frac{\sigma\tilde{t}}{2}(c_{-\sigma}^{\dagger\bar{\alpha}} - c_{-\sigma}^{\dagger\bar{\alpha}'}) DB_S - \frac{\sigma\tilde{t}}{2}(c_{-\sigma}^{\dagger\bar{\alpha}} + c_{-\sigma}^{\dagger\bar{\alpha}'}) DB_A  ,
\end{eqnarray} 

\begin{eqnarray}
i\frac{\partial}{\partial t} FD_S^{\sigma} &=& \left(2\varepsilon_0 - 3\mu +U + \frac{\tilde{t}}{2d}\right) FD_S^{\sigma} - \frac{\sigma\tilde{t}}{\sqrt{2}}(c_{-\sigma}^{\dagger\bar{\alpha}} + c_{-\sigma}^{\dagger\bar{\alpha}'}) DD   
- \frac{\sigma\tilde{t}}{\sqrt{2}}(c_{-\sigma}^{\dagger\bar{\alpha}} - c_{-\sigma}^{\dagger\bar{\alpha}'}) FF^{\sigma} 
\nonumber \\
 &+& \frac{\sigma\tilde{t}}{2}(c_{-\sigma}^{\bar{\alpha}} - c_{-\sigma}^{\bar{\alpha}'}) FF_S - \frac{\sigma\tilde{t}}{2}(c_{-\sigma}^{\bar{\alpha}} + c_{-\sigma}^{\bar{\alpha}'}) FF_A    \nonumber \\
&-& \frac{\tilde{t}}{2}(c_{-\sigma}^{\bar{\alpha}} + c_{-\sigma}^{\bar{\alpha}'}) DB_S + \frac{\tilde{t}}{2}(c_{-\sigma}^{\bar{\alpha}} - c_{-\sigma}^{\bar{\alpha}'}) DB_A  ,
\end{eqnarray}

\begin{eqnarray}
i\frac{\partial}{\partial t} FD_A^{\sigma} &=& \left(2\varepsilon_0 - 3\mu +U - \frac{\tilde{t}}{2d}\right) FD_A^{\sigma} - \frac{\sigma\tilde{t}}{\sqrt{2}}(c_{-\sigma}^{\dagger\bar{\alpha}} - c_{-\sigma}^{\dagger\bar{\alpha}'}) DD   
+ \frac{\sigma\tilde{t}}{\sqrt{2}}(c_{-\sigma}^{\dagger\bar{\alpha}} + c_{-\sigma}^{\dagger\bar{\alpha}'}) FF^{\sigma}
\nonumber \\
 &-& \frac{\sigma\tilde{t}}{2}(c_{-\sigma}^{\bar{\alpha}} + c_{-\sigma}^{\bar{\alpha}'}) FF_S + \frac{\sigma\tilde{t}}{2}(c_{-\sigma}^{\bar{\alpha}} - c_{-\sigma}^{\bar{\alpha}'}) FF_A    \nonumber \\
&-& \frac{\tilde{t}}{2}(c_{-\sigma}^{\bar{\alpha}} - c_{-\sigma}^{\bar{\alpha}'}) DB_S + \frac{\tilde{t}}{2}(c_{-\sigma}^{\bar{\alpha}} + c_{-\sigma}^{\bar{\alpha}'}) DB_A  ,
\end{eqnarray}

\begin{equation}
i\frac{\partial}{\partial t} BB = 2\varepsilon_0 BB - \frac{\tilde{t}}{\sqrt{2}}\sum_{\tau}(c_{\tau}^{\dagger \bar{\alpha}} + c_{\tau}^{\dagger \bar{\alpha}'}) FB_S^{\tau} - 
\frac{\tilde{t}}{\sqrt{2}}\sum_{\tau}(c_{\tau}^{\dagger \bar{\alpha}} - c_{\tau}^{\dagger \bar{\alpha}'}) FB_A^{\tau} ,
\end{equation}

\begin{eqnarray}
i\frac{\partial}{\partial t} DD &=& 2(\varepsilon_0 -2\mu +U) DD \nonumber \\
 &-& \frac{\tilde{t}}{\sqrt{2}}\sum_{\tau}\tau(c_{-\tau}^{\bar{\alpha}} + c_{-\tau}^{\bar{\alpha}'}) FD_S^{\tau} - 
\frac{\tilde{t}}{\sqrt{2}}\sum_{\tau}\tau(c_{-\tau}^{\bar{\alpha}} - c_{-\tau}^{\bar{\alpha}'}) FD_A^{\tau} ,
\end{eqnarray}

\begin{eqnarray}
i\frac{\partial}{\partial t} FF^{\sigma} &=& 2(\varepsilon_0-\mu)FF^{\sigma}
- \frac{\tilde{t}}{\sqrt{2}} (c_{\sigma}^{\bar{\alpha}} - c_{\sigma}^{\bar{\alpha}'}) FB_S^{\sigma} + \frac{\tilde{t}}{\sqrt{2}} (c_{\sigma}^{\bar{\alpha}} + c_{\sigma}^{\bar{\alpha}'}) FB_A^{\sigma} \nonumber \\
&-& \frac{\sigma\tilde{t}}{\sqrt{2}} (c_{-\sigma}^{\dagger\bar{\alpha}} - c_{-\sigma}^{\dagger\bar{\alpha}'}) FD_S^{\sigma} + \frac{\sigma\tilde{t}}{\sqrt{2}} (c_{-\sigma}^{\dagger\bar{\alpha}} + c_{-\sigma}^{\dagger\bar{\alpha}'}) FD_A^{\sigma} ,
\end{eqnarray}

\begin{eqnarray}
i\frac{\partial}{\partial t} FF_S &=& 2(\varepsilon_0-\mu)FF_S \nonumber \\
&-& \frac{\tilde{t}}{2} \sum_{\tau} (c_{-\tau}^{\bar{\alpha}} - c_{-\tau}^{\bar{\alpha}'}) FB_S^{\tau} + \frac{\tilde{t}}{2} \sum_{\tau} (c_{-\tau}^{\bar{\alpha}} + c_{-\tau}^{\bar{\alpha}'}) FB_A^{\tau} \nonumber \\
 &+& \frac{\tilde{t}}{2} \sum_{\tau} \tau(c_{\tau}^{\dagger\tilde{\alpha}} - c_{\tau}^{\dagger\tilde{\alpha}'}) FD_S^{\tau} - \sum_{\tau} \tau(c_{\tau}^{\dagger\tilde{\alpha}} + c_{\tau}^{\dagger\tilde{\alpha}'}) FD_A^{\tau} ,
\end{eqnarray}

\begin{eqnarray}
i\frac{\partial}{\partial t} FF_A &=& 2(\varepsilon_0-\mu)FF_A + \frac{\sigma\tilde{t}}{d}DB_S  \nonumber \\
&+& \frac{\sigma\tilde{t}}{2} \sum_{\tau} \tau(c_{-\tau}^{\bar{\alpha}} + c_{-\tau}^{\bar{\alpha}'}) FB_S^{\tau} - \frac{\sigma\tilde{t}}{2} \sum_{\tau} \tau(c_{-\tau}^{\bar{\alpha}} - c_{-\tau}^{\bar{\alpha}'}) FB_A^{\tau} \nonumber \\
 &-& \frac{\sigma\tilde{t}}{2} \sum_{\tau} (c_{\tau}^{\dagger\tilde{\alpha}} + c_{\tau}^{\dagger\tilde{\alpha}'}) FD_S^{\tau} + \sum_{\tau} (c_{\tau}^{\dagger\tilde{\alpha}} - c_{\tau}^{\dagger\tilde{\alpha}'}) FD_A^{\tau} ,
\end{eqnarray}

\begin{eqnarray}
i\frac{\partial}{\partial t} DB_S &=& (2\varepsilon_0 - 2\mu +U) DB_S + \frac{\tilde{t}}{d} FF_A  \nonumber \\
&-& \frac{\tilde{t}}{2} \sum_{\tau} \tau (c_{-\tau}^{\bar{\alpha}} + c_{-\tau}^{\bar{\alpha}'}) FB_S^{\tau} - \frac{\tilde{t}}{2} \sum_{\tau} \tau (c_{-\tau}^{\bar{\alpha}} - c_{-\tau}^{\bar{\alpha}'}) FB_A^{\tau} \nonumber \\
&-& \frac{\tilde{t}}{2} \sum_{\tau} (c_{\tau}^{\dagger\bar{\alpha}} + c_{\tau}^{\dagger\bar{\alpha}'}) FD_S^{\tau} - \frac{\tilde{t}}{2} \sum_{\tau} (c_{\tau}^{\dagger\bar{\alpha}} - c_{\tau}^{\dagger\bar{\alpha}'}) FD_A^{\tau} ,
\end{eqnarray}

\begin{eqnarray}
i\frac{\partial}{\partial t} DB_A &=& (2\varepsilon_0 - 2\mu +U) DB_A   \nonumber \\
&-& \frac{\tilde{t}}{2} \sum_{\tau} \tau (c_{-\tau}^{\bar{\alpha}} - c_{-\tau}^{\bar{\alpha}'}) FB_S^{\tau} - \frac{\tilde{t}}{2} \sum_{\tau} \tau (c_{-\tau}^{\bar{\alpha}} + c_{-\tau}^{\bar{\alpha}'}) FB_A^{\tau} \nonumber \\
&+& \frac{\tilde{t}}{2} \sum_{\tau} (c_{\tau}^{\dagger\bar{\alpha}} - c_{\tau}^{\dagger\bar{\alpha}'}) FD_S^{\tau} + \frac{\tilde{t}}{2} \sum_{\tau} (c_{\tau}^{\dagger\bar{\alpha}} + c_{\tau}^{\dagger\bar{\alpha}'}) FD_A^{\tau}.
\end{eqnarray}

\section{Appendix B: Self-energies for the two-site resolvents}
The resolvents associated with the two-site operators $\Phi_n$ can be expressed in the form
\begin{equation}
R_{\Phi_n \Phi_m}(\omega) = \left(\omega - E_{\Phi_n \Phi_m} - \Sigma_{\Phi_n \Phi_m}(\omega)\right)^{-1}.
\end{equation}
For the diagonal components, we use the notation $X_{\Phi_n \Phi_m} \equiv X_{\Phi_n}$, where $X$ is $R$, $E$, or $\Sigma$.
We evaluate the self-energies $\Sigma_{\Phi_n \Phi_m}(\omega)$ within a one loop approximation. Using the spectral functions $\rho_+$ and $\rho_-$ , Eq. (\ref{irred}), of the irreducible propagators , we obtain 
\begin{eqnarray} 
\Sigma_{FB_S}(\omega) &=& \frac{\tilde{t}^2}{4} \int dx~ [2\rho_+(x)(1-f(x)) R_{BB}(\omega-x)  - \rho_+(x)f(x)R_{FF_A DB_S} (\omega+x) \nonumber \\
 &+& ~\rho_+(x)f(x)R_{DB_S}(\omega+x) + \rho_-(x)f(x)R_{DB_A}(\omega+x) \nonumber\\
 &+& 3\rho_-(x)f(x)R_{FF_S}(\omega+x) + \rho_+(x)f(x)R_{FF_A}(\omega+x)],
\end{eqnarray}

\begin{eqnarray} 
\Sigma_{FB_A}(\omega) &=& \frac{\tilde{t}^2}{4} \int dx~[2\rho_-(x)(1-f(x)) R_{BB}(\omega-x)  + \rho_-(x)f(x)R_{FF_A DB_S} (\omega+x) \nonumber \\
 &+& ~\rho_-(x)f(x)R_{DB_S}(\omega+x) + \rho_+(x)f(x)R_{DB_A}(\omega+x) \nonumber\\
 &+& 3\rho_+(x)f(x)R_{FF_S}(\omega+x) + \rho_-(x)f(x)R_{FF_A}(\omega+x)] ,
\end{eqnarray}

\begin{eqnarray} 
\Sigma_{FD_S}(\omega) &=& \frac{\tilde{t}^2}{4} \int dx~[2\rho_+(x)f(x) R_{DD}(\omega+x)  + \rho_+(x)(1-f(x))R_{FF_A DB_S} (\omega-x) \nonumber \\
 &+& ~\rho_+(x)(1-f(x))R_{DB_S}(\omega-x) + \rho_-(x)(1-f(x))R_{DB_A}(\omega-x) \nonumber\\
 &+& 3\rho_-(x)(1-f(x))R_{FF_S}(\omega-x) + \rho_+(x)(1-f(x))R_{FF_A}(\omega-x)]  ,
\end{eqnarray}

\begin{eqnarray} 
\Sigma_{FD_A}(\omega) &=& \frac{\tilde{t}^2}{4} \int dx~[2\rho_-(x)f(x) R_{DD}(\omega+x)  - \rho_-(x)(1-f(x))R_{FF_A DB_S} (\omega-x) \nonumber \\
 &+& ~\rho_-(x)(1-f(x))R_{DB_S}(\omega-x) + \rho_+(x)(1-f(x))R_{DB_A}(\omega-x) \nonumber\\
 &+& 3\rho_+(x)(1-f(x))R_{FF_S}(\omega-x) + \rho_-(x)(1-f(x))R_{FF_A}(\omega-x)]  ,
\end{eqnarray}

\begin{equation} 
\Sigma_{BB}(\omega) = \tilde{t}^2 \int dx~[\rho_+(x)f(x)R_{FB_S}(\omega+x)
+  \rho_-(x)f(x)R_{FB_A}(\omega+x)] ,
\end{equation}

\begin{equation} 
\Sigma_{DD}(\omega) = \tilde{t}^2 \int dx~[\rho_+(x)(1-f(x))R_{FD_S}(\omega-x)
+  \rho_-(x)(1-f(x))R_{FD_A}(\omega-x)] ,
\end{equation}

\begin{eqnarray} 
\Sigma_{FF_S}(\omega) &=& \frac{\tilde{t}^2}{2} \int dx~[\rho_-(x)f(x)R_{FD_S}(\omega+x) + \rho_{+}(x)f(x)R_{FD_A}(\omega+x)  \nonumber \\
&+& \rho_-(x)(1-f(x))R_{FB_S}(\omega-x) \nonumber \\
&+& \rho_+(x)(1-f(x))R_{FB_A}(\omega+x)] ,
\end{eqnarray}

\begin{equation} 
\Sigma_{FF^{\sigma}}(\omega) =  \Sigma_{FF_S}(\omega) = \Sigma_{DB_A}(\omega)
\end{equation}

\begin{eqnarray} 
\Sigma_{FF_A}(\omega) &=& \frac{\tilde{t}^2}{2} \int dx~[\rho_+(x)f(x)R_{FD_S}(\omega+x) + \rho_{-}(x)f(x)R_{FD_A}(\omega+x)  \nonumber \\
&+& \rho_+(x)(1-f(x))R_{FB_S}(\omega-x) \nonumber \\
&+& \rho_-(x)(1-f(x))R_{FB_A}(\omega+x)] ,
\end{eqnarray}

\begin{eqnarray} 
\Sigma_{FF_ADB_S}(\omega) &=& \frac{\tilde{t}^2}{2} \int dx~[\rho_+(x)f(x)R_{FD_S}(\omega+x) - \rho_{-}(x)f(x)R_{FD_A}(\omega+x)  \nonumber \\
&-& \rho_+(x)(1-f(x))R_{FB_S}(\omega-x) \nonumber \\
&+& \rho_-(x)(1-f(x))R_{FB_A}(\omega+x)] ,
\end{eqnarray}

\begin{equation} 
\Sigma_{FF_A DB_S}(\omega) = \Sigma_{DB_S FF_A}(\omega) ,
\end{equation}

\begin{equation} 
\Sigma_{DB_S}(\omega) = \Sigma_{FF_A}(\omega) .
\end{equation}

\section{Appendix C: Self-energies corrections due to spin fluctuation}

We include the effects of spin fluctuation as higher order corrections to the
 self-energies of the resolvents. The spectral functions associated with
 local singlet and triplet states are sharply peaked at well defined
 energies separated by the effective antiferromagnetic coupling constant
\begin{equation}
J = \int_{-\infty}^{\infty} \omega (\sigma_{FF_S} - \sigma_{FF_A}) ~d\omega.
\end{equation}
The effects of spin fluctuations with the environment can be approximately described by the
 effective antiferromagnetic interaction
\begin{equation}
\delta H_{eff} = \frac{1}{2} J \vec{n} \vec{n}^\alpha.
\end{equation}
where $\vec{n} = c^\dagger \vec{\sigma} c$ is the spin density, $\sigma_i$
 are the Pauli matrices and $c^{\dagger} = (c_{\uparrow}^{\dagger},
 c_{\downarrow}^{\dagger})$. 
Although a proper account of the
singlet-triplet mixing can be given by considering vertex corrections,
we can use a simpler two-step approach: 1) Working in the basis formed
by the eigenstates of the two-site problem (as we, in fact, do) we take
care of the spin fluctuations for the cluster without a bath. As we
introduce the bath, an additional singlet-triplet mixing occurs which is
not captured by NCA (as shown by the very sharp features in the FFA and
the FFS resolvents). We try to approximate  this mixing by an effective
spin-spin interaction.  As a result of this mixing,  the FFA and FFS states are broadened. 2) Due to the  the self-consistency of the approach,
this spin-spin interaction is also present between the cluster and the
bath. The corrections that we introduce are the effect of this additional
effective interaction with the bath on the self-energies of the resolvents.
 Consequently, some of the
 equations of motion for the two-site level operators
 $\Phi_n$ (see Appendix A), will be modified. Explicitly we have:
\begin{eqnarray}
i\frac{\partial}{\partial t}FB_S^{\sigma} &=& \frac{J}{2}
 [\sigma(n_3^{\bar{\alpha}} + n_3^{\bar{\alpha}'})FB_S^{\sigma}
 + \sigma(n_3^{\bar{\alpha}} - n_3^{\bar{\alpha}'})FB_A^{\sigma} 
\nonumber \\
&+& (n_{-\sigma}^{\bar{\alpha}} + n_{-\sigma}^{\bar{\alpha}'})FB_S^{-\sigma} + 
(n_{-\sigma}^{\bar{\alpha}} - n_{-\sigma}^{\bar{\alpha}'})FB_A^{-\sigma}] + \cdots,
\end{eqnarray}

\begin{eqnarray}
i\frac{\partial}{\partial t}FB_A^{\sigma} &=& \frac{J}{2} [\sigma(n_3^{\bar{\alpha}} - n_3^{\bar{\alpha}'})FB_S^{\sigma} + \sigma(n_3^{\bar{\alpha}} + n_3^{\bar{\alpha}'})FB_A^{\sigma} \nonumber \\
&+& (n_{-\sigma}^{\bar{\alpha}} - n_{-\sigma}^{\bar{\alpha}'})FB_S^{-\sigma} + (n_{-\sigma}^{\bar{\alpha}} + n_{-\sigma}^{\bar{\alpha}'})FB_A^{-\sigma}] + \cdots,
\end{eqnarray}

\begin{eqnarray}
i\frac{\partial}{\partial t}FD_S^{\sigma} &=& \frac{J}{2} [\sigma(n_3^{\bar{\alpha}} + n_3^{\bar{\alpha}'})FD_S^{\sigma} + \sigma(n_3^{\bar{\alpha}} - n_3^{\bar{\alpha}'})FD_A^{\sigma} \nonumber \\
&+& (n_{-\sigma}^{\bar{\alpha}} + n_{-\sigma}^{\bar{\alpha}'})FD_S^{-\sigma} + (n_{-\sigma}^{\bar{\alpha}} - n_{-\sigma}^{\bar{\alpha}'})FD_A^{-\sigma}] + \cdots
\end{eqnarray}

\begin{eqnarray}
i\frac{\partial}{\partial t}FD_A^{\sigma} &=& \frac{J}{2} [\sigma(n_3^{\bar{\alpha}} - n_3^{\bar{\alpha}'})FD_S^{\sigma} + \sigma(n_3^{\bar{\alpha}} + n_3^{\bar{\alpha}'})FD_A^{\sigma} \nonumber \\
&+& (n_{-\sigma}^{\bar{\alpha}} - n_{-\sigma}^{\bar{\alpha}'})FD_S^{-\sigma} + (n_{-\sigma}^{\bar{\alpha}} + n_{-\sigma}^{\bar{\alpha}'})FD_A^{-\sigma}] + \cdots,
\end{eqnarray}

\begin{eqnarray}
i\frac{\partial}{\partial t}FF^{\sigma} &=& J [\sigma(n_3^{\bar{\alpha}} + n_3^{\bar{\alpha}'})FF^{\sigma} + \frac{1}{\sqrt{2}}(n_{-\sigma}^{\bar{\alpha}} + n_{-\sigma}^{\bar{\alpha}'}) FF_S   \nonumber \\
&-&  \frac{\sigma}{\sqrt{2}}(n_{-\sigma}^{\bar{\alpha}} - n_{-\sigma}^{\bar{\alpha}'}) FF_A  ] + \cdots,
\end{eqnarray}

\begin{eqnarray}
i\frac{\partial}{\partial t}FF_S &=& J [\sigma(n_3^{\bar{\alpha}} - n_3^{\bar{\alpha}'})FF_A + \frac{1}{\sqrt{2}}(n_{-\sigma}^{\bar{\alpha}} + n_{-\sigma}^{\bar{\alpha}'}) FF^{\sigma}   \nonumber \\
&+&  \frac{1}{\sqrt{2}}(n_{-\sigma}^{\bar{\alpha}} + n_{-\sigma}^{\bar{\alpha}'}) FF^{-\sigma} ] + \cdots,
\end{eqnarray}

\begin{eqnarray}
i\frac{\partial}{\partial t}FF_A &=& J [\sigma(n_3^{\bar{\alpha}} - n_3^{\bar{\alpha}'})FF_S - \frac{1}{\sqrt{2}}(n_{-\sigma}^{\bar{\alpha}} - n_{-\sigma}^{\bar{\alpha}'}) FF^{\sigma}   \nonumber \\
&+&  \frac{1}{\sqrt{2}}(n_{-\sigma}^{\bar{\alpha}} - n_{-\sigma}^{\bar{\alpha}'}) FF^{-\sigma} ] + \cdots,
\end{eqnarray}
where we used the notation $n_{\sigma} = n_1 +i\sigma n_2$. These terms will determine  the self-energy corrections
\begin{eqnarray}
\delta\Sigma_{FB_S}(\omega) &=& \int dx~\frac{3}{2}J^2[\varphi_+(x)(1-f(x))R_{FB_S}(\omega-x) \nonumber \\
&+& \varphi_-(x)(1-f(x))R_{FB_A}(\omega-x)],
\end{eqnarray}

\begin{eqnarray}
\delta\Sigma_{FB_A}(\omega) &=& \int dx~\frac{3}{2}J^2[\varphi_-(x)(1-f(x))R_{FB_S}(\omega-x) \nonumber \\
&+& \varphi_+(x)(1-f(x))R_{FB_A}(\omega-x)],
\end{eqnarray}

\begin{eqnarray}
\delta\Sigma_{FD_S}(\omega) &=& \int dx~\frac{3}{2}J^2[\varphi_+(x)(1-f(x))R_{FD_S}(\omega-x) \nonumber \\
&+& \varphi_-(x)(1-f(x))R_{FD_A}(\omega-x)],
\end{eqnarray}

\begin{eqnarray}
\delta\Sigma_{FD_A}(\omega) &=& \int dx~\frac{3}{2}J^2[\varphi_-(x)(1-f(x))R_{FD_S}(\omega-x) \nonumber \\
&+& \varphi_+(x)(1-f(x))R_{FD_A}(\omega-x)],
\end{eqnarray}

\begin{eqnarray}
\delta\Sigma_{FF_S}(\omega) &=& \int dx~2J^2[2\varphi_+(x)(1-f(x))R_{FF_S}(\omega-x) \nonumber \\
&+& \varphi_-(x)(1-f(x))R_{FF_A}(\omega-x)],
\end{eqnarray}

\begin{equation}
\delta\Sigma_{FF_A}(\omega) = \int dx~6J^2 \varphi_-(x)(1-f(x))R_{FF_S}(\omega-x),
\end{equation}
where $\varphi_{\pm}(\omega)  = \varphi(\omega) \pm \varphi '(\omega)$ and 
\begin{eqnarray}
\varphi(\omega) &=& F.T.\langle\{n_3^{\bar{\alpha}}, n_3^{\bar{\alpha}}\}\rangle = \frac{1}{2} F.T.\langle\{n_{-\sigma}^{\bar{\alpha}}, n_{-\sigma}^{\bar{\alpha}}\}\rangle  \nonumber \\
\varphi '(\omega) &=& F.T.\langle\{n_3^{\bar{\alpha}}, n_3^{\bar{\alpha}'}\}\rangle = \frac{1}{2} F.T.\langle\{n_{-\sigma}^{\bar{\alpha}}, n_{-\sigma}^{\bar{\alpha}'}\}\rangle 
\end{eqnarray}
Note because $\varphi(\omega)$ and $\varphi'(\omega)$ are expressed in terms of functions defined on $\bar{\alpha}$ and $\bar{\alpha'}$ which denote the nearest neighbours of the two sites in the cluster, we have avoided double counting the spin fluctuations on the cluster.
The functions $\varphi(\omega)$ and $\varphi '(\omega)$ are computed within a non-crossing approximation, taking into account the only dominant contributions (containing products of on-site and nearest-neighbor operators). Introducing the coordination factors
$\lambda = (2d-1)/((2d)^2)$ and $\lambda ' = (2d-2)/((2d)^2)$, we obtain

\begin{eqnarray}
\varphi(\omega) &\approx& \frac{\lambda}{Z} \int dx \left\{\frac{1}{2}[\sigma_{FB_S}(x)\bar{\sigma}_{FB_S}(\omega -x) + \bar{\sigma}_{FB_S}(x)\sigma_{FB_S}(\omega -x)]\right. \nonumber \\
&~~~~~~~+& \frac{1}{2}[\sigma_{FB_S}(x)\bar{\sigma}_{FB_A}(\omega -x) + \bar{\sigma}_{FB_S}(x)\sigma_{FB_A}(\omega -x)] \nonumber \\
&~~~~~~~+& \frac{1}{2}[\sigma_{FB_A}(x)\bar{\sigma}_{FB_S}(\omega -x) + \bar{\sigma}_{FB_A}(x)\sigma_{FB_S}(\omega -x)] \nonumber \\
&~~~~~~~+& \frac{1}{2}[\sigma_{FB_A}(x)\bar{\sigma}_{FB_A}(\omega -x) + \bar{\sigma}_{FB_A}(x)\sigma_{FB_A}(\omega -x)]  \nonumber \\
&~~~~~~~+& \frac{1}{2}[\sigma_{FD_S}(x)\bar{\sigma}_{FD_S}(\omega -x) + \bar{\sigma}_{FB_S}(x)\sigma_{FB_S}(\omega -x)] \nonumber \\
&~~~~~~~+& \frac{1}{2}[\sigma_{FD_S}(x)\bar{\sigma}_{FD_A}(\omega -x) + \bar{\sigma}_{FB_S}(x)\sigma_{FB_A}(\omega -x)] \nonumber \\
&~~~~~~~+& \frac{1}{2}[\sigma_{FD_A}(x)\bar{\sigma}_{FD_S}(\omega -x) + \bar{\sigma}_{FB_A}(x)\sigma_{FB_S}(\omega -x)] \nonumber \\
&~~~~~~~+& \frac{1}{2}[\sigma_{FD_A}(x)\bar{\sigma}_{FD_A}(\omega -x) + \bar{\sigma}_{FB_A}(x)\sigma_{FB_A}(\omega -x)]  \nonumber \\
&~~~~~~~+& 2[\sigma_{FF_S}(x)\bar{\sigma}_{FF_S}(\omega -x) + \bar{\sigma}_{FF_S}(x)\sigma_{FF_S}(\omega -x)]  \nonumber \\
&~~~~~~~+& ~~\sigma_{FF_S}(x)\bar{\sigma}_{FF_A}(\omega -x) + \bar{\sigma}_{FF_S}(x)\sigma_{FF_A}(\omega -x)  \nonumber \\
&~~~~~~~+& ~\left.\frac{}{}\sigma_{FF_A}(x)\bar{\sigma}_{FF_S}(\omega -x) + \bar{\sigma}_{FF_A}(x)\sigma_{FF_S}(\omega -x)\right\} ,
\end{eqnarray}
\begin{eqnarray}
\varphi '(\omega) &\approx& \frac{\lambda '}{Z} \int dx \left\{\frac{1}{2}[\sigma_{FB_S}(x)\bar{\sigma}_{FB_S}(\omega -x) + \bar{\sigma}_{FB_S}(x)\sigma_{FB_S}(\omega -x)]\right. \nonumber \\
&~~~~~~~-& \frac{1}{2}[\sigma_{FB_S}(x)\bar{\sigma}_{FB_A}(\omega -x) - \bar{\sigma}_{FB_S}(x)\sigma_{FB_A}(\omega -x)] \nonumber \\
&~~~~~~~-& \frac{1}{2}[\sigma_{FB_A}(x)\bar{\sigma}_{FB_S}(\omega -x) - \bar{\sigma}_{FB_A}(x)\sigma_{FB_S}(\omega -x)] \nonumber \\
&~~~~~~~+& \frac{1}{2}[\sigma_{FB_A}(x)\bar{\sigma}_{FB_A}(\omega -x) + \bar{\sigma}_{FB_A}(x)\sigma_{FB_A}(\omega -x)]  \nonumber \\
&~~~~~~~+& \frac{1}{2}[\sigma_{FD_S}(x)\bar{\sigma}_{FD_S}(\omega -x) + \bar{\sigma}_{FB_S}(x)\sigma_{FB_S}(\omega -x)] \nonumber \\
&~~~~~~~-& \frac{1}{2}[\sigma_{FD_S}(x)\bar{\sigma}_{FD_A}(\omega -x) - \bar{\sigma}_{FB_S}(x)\sigma_{FB_A}(\omega -x)] \nonumber \\
&~~~~~~~-& \frac{1}{2}[\sigma_{FD_A}(x)\bar{\sigma}_{FD_S}(\omega -x) - \bar{\sigma}_{FB_A}(x)\sigma_{FB_S}(\omega -x)] \nonumber \\
&~~~~~~~+& \frac{1}{2}[\sigma_{FD_A}(x)\bar{\sigma}_{FD_A}(\omega -x) + \bar{\sigma}_{FB_A}(x)\sigma_{FB_A}(\omega -x)]  \nonumber \\
&~~~~~~~+& 2[\sigma_{FF_S}(x)\bar{\sigma}_{FF_S}(\omega -x) + \bar{\sigma}_{FF_S}(x)\sigma_{FF_S}(\omega -x)]  \nonumber \\
&~~~~~~~-& ~~\sigma_{FF_S}(x)\bar{\sigma}_{FF_A}(\omega -x) - \bar{\sigma}_{FF_S}(x)\sigma_{FF_A}(\omega -x)  \nonumber \\
&~~~~~~~-& ~\left.\frac{}{}\sigma_{FF_A}(x)\bar{\sigma}_{FF_S}(\omega -x) - \bar{\sigma}_{FF_A}(x)\sigma_{FF_S}(\omega -x)\right\}  .
\end{eqnarray}
Note that in these equations, it is the Fermi function which appears not the Bose distribution function.  In general, the expression for the self energy depends on which form one uses for the spectral function.  We used
$F.T.\angle\{n_3(t),n_3(t')\}\langle=\varphi(\omega)$. Usually the
pectral function is defined by
$F.T.\langle[n_3(t),n_3(t')]\rangle=\rho(\omega)$. They are related as
$\varphi(\omega)=[(e^{\beta\omega}+1)/(e^{\beta\omega}-1)]\rho(\omega)$.
The loop contains
\beq
\int d\omega'\rho(\omega'+\omega)R(-\omega')
 \frac{e^{\beta(\omega+\omega')}}{e^{\beta(\omega+\omega')}-1}
\eeq
which, in terms of $ \varphi$, becomes
\beq
\int d\omega'\varphi(\omega'+\omega)R(-\omega')
 \frac{e^{\beta(\omega+\omega')}}{e^{\beta(\omega+\omega')}+1},
\eeq
thereby justifying the Fermi function in the self-energy corrections.

\section{Appendix D:Level operator representation of $\delta J$ and explicit expressions for $Dm_0$ and $Dm_1$}

With the assumption used to derive Eq. (\ref{DmEq}), we find
that the the dynamical correction operator $\delta J$, Eq. (\ref{delphi}), 
 can be expressed in terms of two-site level operators,
\begin{equation}
\delta J_{\sigma} = -\tilde{t}\sum_{m,n} a_{nm} \Phi_n^{\dagger}\Phi_m.
\end{equation}
With the notation
\begin{eqnarray}
a_1 &=& eI_1^{-1} ,~~~~~~~~~~~~~~~~~b_1 = \frac{n}{2} - \tilde{p}I_1^{-1}, \nonumber\\
a_2 &=& eI_2^{-1} ,~~~~~~~~~~~~~~~~~b_2 = \frac{n}{2} + \tilde{p}I_2^{-1},
\end{eqnarray}
we obtain for the coefficients $a_{nm}$:
\begin{eqnarray}
a_{BB~FB_S^{\sigma}} &=& \frac{-1}{\sqrt{2}}(a_1-b_1), ~~~
a_{FD_S^{-\sigma}~DD} = \frac{-\sigma}{\sqrt{2}} (1-a_2-b_2), \nonumber\\
a_{BB~FB_A^{\sigma}} &=& \frac{-1}{\sqrt{2}}(a_1+b_1), ~~~
a_{FD_A^{-\sigma}~DD}  =  \frac{\sigma}{\sqrt{2}} (1+a_2-b_2),\nonumber\\
a_{FB_S^{-\sigma}~FF_S} &=& \frac{1}{2}(a_1+b_1),  ~~~~~~
a_{FF_S~FD_S^{\sigma}} = \frac{-\sigma}{2} (1+a_2-b_2), \nonumber\\
a_{FB_A^{-\sigma}~FF_S} &=& \frac{-1}{2}(a_1-b_1),  ~~~~
a_{FF_S~FD_A^{\sigma}} = \frac{-\sigma}{2} (1-a_2-b_2), \nonumber
\end{eqnarray}

\begin{eqnarray}
a_{FB_S^{\sigma}~FF^{\sigma}} &=& \frac{1}{\sqrt{2}}(a_1+b_1),  ~~~~~
a_{FF^{-\sigma}~FD_S^{-\sigma}} = \frac{-\sigma}{\sqrt{2}} (1+a_2-b_2), \nonumber\\
a_{FB_A^{\sigma}~FF^{\sigma}} &=& \frac{-1}{\sqrt{2}}(a_1-b_1),  ~~~~~
a_{FF^{-\sigma}~FD_A^{-\sigma}} = \frac{-\sigma}{\sqrt{2}} (1-a_2-b_2), \nonumber\\
a_{FB_S^{-\sigma}~FF_A} &=& \sigma (1 +a_1-b_1),~~~~~~
a_{FF_A~FD_S^{\sigma}} = \frac{1}{2} (1+a_2+b_2), \nonumber\\
a_{FB_A^{-\sigma}~FF_A} &=& \sigma (1 -a_1-b_1), ~~~~~~
a_{FF_A~FD_A^{\sigma}} = \frac{1}{2} (1-a_2+b_2), \\
a_{DB_S~FD_S^{\sigma}} &=& \frac{-1}{2}(2+a_1-b_1) ,~~
a_{FB_S^{-\sigma}~DB_S} = \frac{\sigma}{2}(1+a_2+b_2), \nonumber\\
a_{DB_S~FD_A^{\sigma}} &=& ~\frac{1}{2}(2-a_1-b_1) ,~~~
a_{FB_A^{-\sigma}~DB_S} = \frac{-\sigma}{2}(1-a_2+b_2), \nonumber\\
a_{DB_A~FD_S^{\sigma}} &=& ~\frac{1}{2}(a_1+b_1), ~~~~~~~~
a_{FB_S^{-\sigma}~DB_A} = \frac{\sigma}{2}(1+a_2-b_2), \nonumber\\
a_{DB_A~FD_A^{\sigma}} &=& ~\frac{1}{2}(a_1-b_1), ~~~~~~~~
a_{FB_A^{-\sigma}~DB_A} = \frac{-\sigma}{2}(1-a_2-b_2). \nonumber
\end{eqnarray}
Using these coefficients, we can write explicitly the dynamical corrections $Dm_0$ and $Dm_1$. From Eq. (\ref{DmEq}), we have
\begin{eqnarray}
Dm_0 (\omega) &=& \frac{\tilde{t}^2}{Z}\int dxdx'~\frac{1}{\omega-x+x'+i\delta} \\
&\times& \left\{a_{BB~FB_S}^2[\sigma_{FB_S}(x)\bar{\sigma}_{BB}(x') + \bar{\sigma}_{FB_S}(x)\sigma_{BB}(x')]\frac{}{}\right. \nonumber\\
&~~+& ~~a_{BB~FB_A}^2[\sigma_{FB_A}(x)\bar{\sigma}_{BB}(x') + \bar{\sigma}_{FB_A}(x)\sigma_{BB}(x')] \nonumber\\
&~~+& ~~a_{FD_S~DD}^2[\sigma_{DD}(x)\bar{\sigma}_{FD_S}(x') + \bar{\sigma}_{DD}(x)\sigma_{FD_S}(x')] \nonumber\\
&~~+& ~~a_{FD_A~DD}^2[\sigma_{DD}(x)\bar{\sigma}_{FD_A}(x') + \bar{\sigma}_{DD}(x)\sigma_{FD_A}(x')] \nonumber\\
&~~+& ~(a_{FB_S~FF_S}^2 + a_{FB_S~FF^{\sigma}}^2)[\sigma_{FF_S}(x)\bar{\sigma}_{FB_S}(x') + \bar{\sigma}_{FF_S}(x)\sigma_{FB_S}(x')] \nonumber\\
&~~+& ~(a_{FB_A~FF_S}^2 + a_{FB_A~FF^{\sigma}}^2)[\sigma_{FF_S}(x)\bar{\sigma}_{FB_A}(x') + \bar{\sigma}_{FF_S}(x)\sigma_{FB_A}(x')] \nonumber\\
&~~+& ~(a_{FF_S~FD_S}^2 + a_{FF^{\sigma}~FD_S}^2)[\sigma_{FD_S}(x)\bar{\sigma}_{FF_S}(x') + \bar{\sigma}_{FD_S}(x)\sigma_{FF_S}(x')] \nonumber\\
&~~+& ~(a_{FF_S~FD_A}^2 + a_{FF^{\sigma}~FD_A}^2)[\sigma_{FD_A}(x)\bar{\sigma}_{FF_S}(x') + \bar{\sigma}_{FD_A}(x)\sigma_{FF_S}(x')] \nonumber \\
&~~+& ~~a_{FB_S~FF_A}^2 [\sigma_{FF_A}(x)\bar{\sigma}_{FB_S}(x') + \bar{\sigma}_{FF_A}(x)\sigma_{FB_S}(x')] \nonumber\\
&~~+& ~~a_{FB_A~FF_A}^2 [\sigma_{FF_A}(x)\bar{\sigma}_{FB_A}(x') + \bar{\sigma}_{FF_A}(x)\sigma_{FB_A}(x')] \nonumber\\
&~~+& ~~a_{FF_A~FD_S}^2 [\sigma_{FD_S}(x)\bar{\sigma}_{FF_A}(x') + \bar{\sigma}_{FD_S}(x)\sigma_{FF_A}(x')] \nonumber\\
&~~+& ~~a_{FF_A~FD_A}^2 [\sigma_{FD_A}(x)\bar{\sigma}_{FF_A}(x') + \bar{\sigma}_{FD_A}(x)\sigma_{FF_A}(x')] \nonumber
\end{eqnarray}

\begin{eqnarray}
~~~~~~~~~~~~~&+& ~2a_{FB_S~DB_S}a_{FB_S~FF_A}[\sigma_{DB_S~FF_A}(x)\bar{\sigma}_{FB_S}(x') + \bar{\sigma}_{DB_S~FF_A}(x)\sigma_{FB_S}(x')]  \nonumber\\
&+& ~2a_{FB_A~DB_S}a_{FB_A~FF_A}[\sigma_{DB_S~FF_A}(x)\bar{\sigma}_{FB_A}(x') + \bar{\sigma}_{DB_S~FF_A}(x)\sigma_{FB_A}(x')]  \nonumber\\
&+& ~2a_{DB_S~FD_S}a_{FF_A~FD_S}[\sigma_{FD_S}(x)\bar{\sigma}_{DB_S~FF_A}(x') + \bar{\sigma}_{FD_S}(x)\sigma_{DB_S~FF_A}(x') ]\nonumber\\
&+& ~2a_{DB_S~FD_A}a_{FF_A~FD_A}[\sigma_{FD_A}(x)\bar{\sigma}_{DB_S~FF_A}(x') + \bar{\sigma}_{FD_A}(x)\sigma_{DB_S~FF_A}(x')] \nonumber\\
&+& ~a_{FB_S~DB_S}^2 [\sigma_{DB_S}(x)\bar{\sigma}_{FB_S}(x') + \bar{\sigma}_{DB_S}(x)\sigma_{FB_S}(x')] \nonumber\\
&+& ~a_{FB_A~DB_S}^2 [\sigma_{DB_S}(x)\bar{\sigma}_{FB_A}(x') + \bar{\sigma}_{DB_S}(x)\sigma_{FB_A}(x')] \nonumber\\
&+& ~a_{DB_S~FD_S}^2 [\sigma_{FD_S}(x)\bar{\sigma}_{DB_S}(x') + \bar{\sigma}_{FD_S}(x)\sigma_{DB_S}(x')] \nonumber\\
&+& ~a_{DB_S~FD_A}^2 [\sigma_{FD_A}(x)\bar{\sigma}_{DB_S}(x') + \bar{\sigma}_{FD_A}(x)\sigma_{DB_S}(x')] \nonumber\\
&+& ~a_{FB_S~DB_A}^2 [\sigma_{DB_A}(x)\bar{\sigma}_{FB_S}(x') + \bar{\sigma}_{DB_A}(x)\sigma_{FB_S}(x')] \nonumber\\
&+& ~a_{FB_A~DB_A}^2 [\sigma_{DB_A}(x)\bar{\sigma}_{FB_A}(x') + \bar{\sigma}_{DB_A}(x)\sigma_{FB_A}(x')] \nonumber\\
&+& ~a_{DB_A~FD_S}^2 [\sigma_{FD_S}(x)\bar{\sigma}_{DB_A}(x') + \bar{\sigma}_{FD_S}(x)\sigma_{DB_A}(x')] \nonumber\\
&+& \left.\frac{}{}a_{DB_A~FD_A}^2 [\sigma_{FD_A}(x)\bar{\sigma}_{DB_A}(x') + \bar{\sigma}_{FD_A}(x)\sigma_{DB_A}(x')]\right\}.  \nonumber
\end{eqnarray}
Similarly, for $Dm_1$ we obtain
\begin{eqnarray}
Dm_1 (\omega) &=& \frac{\tilde{t}^2}{Z}\int dxdx'~\frac{1}{\omega-x+x'+i\delta} \\
&\times& \left\{a_{BB~FB_S}^2[\sigma_{FB_S}(x)\bar{\sigma}_{BB}(x') + \bar{\sigma}_{FB_S}(x)\sigma_{BB}(x')]\frac{}{}\right. \nonumber\\
&~~-& ~~a_{BB~FB_A}^2[\sigma_{FB_A}(x)\bar{\sigma}_{BB}(x') + \bar{\sigma}_{FB_A}(x)\sigma_{BB}(x')] \nonumber\\
&~~+& ~~a_{FD_S~DD}^2[\sigma_{DD}(x)\bar{\sigma}_{FD_S}(x') + \bar{\sigma}_{DD}(x)\sigma_{FD_S}(x')] \nonumber\\
&~~-& ~~a_{FD_A~DD}^2[\sigma_{DD}(x)\bar{\sigma}_{FD_A}(x') + \bar{\sigma}_{DD}(x)\sigma_{FD_A}(x')] \nonumber\\
&~~-& ~(a_{FB_S~FF_S}^2 + a_{FB_S~FF^{\sigma}}^2)[\sigma_{FF_S}(x)\bar{\sigma}_{FB_S}(x') + \bar{\sigma}_{FF_S}(x)\sigma_{FB_S}(x')] \nonumber\\
&~~+& ~(a_{FB_A~FF_S}^2 + a_{FB_A~FF^{\sigma}}^2)[\sigma_{FF_S}(x)\bar{\sigma}_{FB_A}(x') + \bar{\sigma}_{FF_S}(x)\sigma_{FB_A}(x')] \nonumber\\
&~~-& ~(a_{FF_S~FD_S}^2 + a_{FF^{\sigma}~FD_S}^2)[\sigma_{FD_S}(x)\bar{\sigma}_{FF_S}(x') + \bar{\sigma}_{FD_S}(x)\sigma_{FF_S}(x')] \nonumber\\
&~~+& ~(a_{FF_S~FD_A}^2 + a_{FF^{\sigma}~FD_A}^2)[\sigma_{FD_A}(x)\bar{\sigma}_{FF_S}(x') + \bar{\sigma}_{FD_A}(x)\sigma_{FF_S}(x')] \nonumber 
\end{eqnarray}

\begin{eqnarray}
~~~~~~~~~~~~~&+&~ a_{FB_S~FF_A}^2 [\sigma_{FF_A}(x)\bar{\sigma}_{FB_S}(x') + \bar{\sigma}_{FF_A}(x)\sigma_{FB_S}(x')] \nonumber\\
&-& ~a_{FB_A~FF_A}^2 [\sigma_{FF_A}(x)\bar{\sigma}_{FB_A}(x') + \bar{\sigma}_{FF_A}(x)\sigma_{FB_A}(x')] \nonumber\\
&+& ~a_{FF_A~FD_S}^2 [\sigma_{FD_S}(x)\bar{\sigma}_{FF_A}(x') + \bar{\sigma}_{FD_S}(x)\sigma_{FF_A}(x')] \nonumber\\
&-& ~a_{FF_A~FD_A}^2 [\sigma_{FD_A}(x)\bar{\sigma}_{FF_A}(x') + \bar{\sigma}_{FD_A}(x)\sigma_{FF_A}(x')] \nonumber\\
&+& ~2a_{FB_S~DB_S}a_{FB_S~FF_A}[\sigma_{DB_S~FF_A}(x)\bar{\sigma}_{FB_S}(x') + \bar{\sigma}_{DB_S~FF_A}(x)\sigma_{FB_S}(x')]  \nonumber\\
&-& ~2a_{FB_A~DB_S}a_{FB_A~FF_A}[\sigma_{DB_S~FF_A}(x)\bar{\sigma}_{FB_A}(x') + \bar{\sigma}_{DB_S~FF_A}(x)\sigma_{FB_A}(x')]  \nonumber\\
&+& ~2a_{DB_S~FD_S}a_{FF_A~FD_S}[\sigma_{FD_S}(x)\bar{\sigma}_{DB_S~FF_A}(x') + \bar{\sigma}_{FD_S}(x)\sigma_{DB_S~FF_A}(x') ]\nonumber\\
&-& ~2a_{DB_S~FD_A}a_{FF_A~FD_A}[\sigma_{FD_A}(x)\bar{\sigma}_{DB_S~FF_A}(x') + \bar{\sigma}_{FD_A}(x)\sigma_{DB_S~FF_A}(x')] \nonumber\\
&+& ~a_{FB_S~DB_S}^2 [\sigma_{DB_S}(x)\bar{\sigma}_{FB_S}(x') + \bar{\sigma}_{DB_S}(x)\sigma_{FB_S}(x')] \nonumber\\
&-& ~a_{FB_A~DB_S}^2 [\sigma_{DB_S}(x)\bar{\sigma}_{FB_A}(x') + \bar{\sigma}_{DB_S}(x)\sigma_{FB_A}(x')] \nonumber\\
&+& ~a_{DB_S~FD_S}^2 [\sigma_{FD_S}(x)\bar{\sigma}_{DB_S}(x') + \bar{\sigma}_{FD_S}(x)\sigma_{DB_S}(x')] \nonumber\\
&-& ~a_{DB_S~FD_A}^2 [\sigma_{FD_A}(x)\bar{\sigma}_{DB_S}(x') + \bar{\sigma}_{FD_A}(x)\sigma_{DB_S}(x')] \nonumber\\
&-& ~a_{FB_S~DB_A}^2 [\sigma_{DB_A}(x)\bar{\sigma}_{FB_S}(x') + \bar{\sigma}_{DB_A}(x)\sigma_{FB_S}(x')] \nonumber\\
&+& ~a_{FB_A~DB_A}^2 [\sigma_{DB_A}(x)\bar{\sigma}_{FB_A}(x') + \bar{\sigma}_{DB_A}(x)\sigma_{FB_A}(x')] \nonumber\\
&-& ~a_{DB_A~FD_S}^2 [\sigma_{FD_S}(x)\bar{\sigma}_{DB_A}(x') + \bar{\sigma}_{FD_S}(x)\sigma_{DB_A}(x')] \nonumber\\
&+& \left.\frac{}{}.a_{DB_A~FD_A}^2 [\sigma_{FD_A}(x)\bar{\sigma}_{DB_A}(x') + \bar{\sigma}_{FD_A}(x)\sigma_{DB_A}(x')]\right\}.  \nonumber
\end{eqnarray}

\section{Appendix E: Static Approximation}

We show explicitly in this appendix that the Hubbard operator technique
in the static approximation\cite{am} correctly recovers the non-interacting or Fermi liquid limit when
$U=0$ regardless of the filling and as $n\rightarrow 0$ for any $U$. Consequently, the violation of Luttinger's theorem found here is not an artificat of the method. Consider the Hubbard operator basis, $\psi_1=\xi_{i\sigma}$ and $\psi_2=\eta_{i\sigma}$ and the associated Green functions $S_{\alpha \beta} = \langle\langle \psi_{\alpha};\psi_{\beta} \rangle\rangle$. Within the static approximation\cite{am}, the expression for the retarded Green function in Fourier space becomes
\beq
S_{\alpha \beta}({\bf k},{\bf \omega}) 
= \sum_{j=1}^2 \frac{\sigma_{\alpha \beta}^{(j)}({\bf k})}
{\omega - \epsilon_j({\bf k}) + i\delta}.   
\eeq
The dispersion relations for the two bands are given by  
$\epsilon_{1,2}({\bf k}) = R({\bf k})\pm  Q({\bf k})$ 
with
\beq
R({\bf k}) &=& \frac{1}{2}U - \mu - \frac{1}{2 I_1 I_2}[m({\bf k}) 
+ 8t\alpha({\bf k})I_1I_2], \\
Q({\bf k}) &=& \frac{1}{2}\sqrt{g^2({\bf k}) + \frac{4m^2({\bf k})}{I_1I_2}},
\eeq
where $I_1 = 1-n/2$ and $I_2 = n/2$ and $\alpha({\bf k}) = \frac{1}{2}(\cos(k_x) + \cos(k_y))$ and we used the notation
$m({\bf k}) = 4t[e + \alpha({\bf k})(p-I_2)]$ and
$g({\bf k}) = -U + (1-n)/(I_1I_2 m({\bf k}))$
where $e=\langle\xi^\dagger_{i\sigma}\xi^\alpha_{i\sigma}\rangle -
\langle\eta_{i\sigma}^\alpha\eta_{i\sigma}\rangle$
and $p=\langle n_{i\sigma}n^\alpha_{i\sigma}\rangle+\langle S_iS_i^{\dagger\alpha}\rangle-\langle b_ib_i^{\dagger\alpha}\rangle$ with $S_i=c_{i\downarrow}^\dagger c_{i\uparrow}$, $b_i=c_{i\uparrow} c_{i\downarrow}$ and $\alpha$ indicates a sum over nearest neighbours of site $i$.
The explicit  expressions for the spectral functions $\sigma_{\alpha \beta}^{(j)}({\bf k})$ are given by 
\beq
\sigma_{\xi\xi}^{(1)}({\bf k}) &=& \frac{I_1}{2} \left[1 + \frac{g({\bf k})}{2Q({\bf k})}\right], ~~~~~ 
\sigma_{\xi\xi}^{(2)}({\bf k}) ~=~ \frac{I_1}{2} \left[1 - \frac{g({\bf k})}{2Q({\bf k})}\right],   \nonumber \\
\sigma_{\xi\eta}^{(1)}({\bf k}) &=& \frac{m({\bf k})}{2Q ({\bf k})}, ~~~~~~~~~~~~~~~
\sigma_{\xi\eta}^{(2)}({\bf k}) ~=~ -\frac{m({\bf k})}{2Q ({\bf k})}, \\
\sigma_{\eta\eta}^{(1)}({\bf k}) &=& \frac{I_2}{2} \left[1 - \frac{g({\bf k})}{2Q({\bf k})}\right],  ~~~~~    
\sigma_{\eta\eta}^{(2)}({\bf k}) ~=~ \frac{I_2}{2} \left[1 + \frac{g({\bf k})}{2Q({\bf k})}\right],   \nonumber
\eeq
Note the fact that the spectral functions are ${\bf k}$-dependent and they also depend on the doping level, through $I_1$ and $I_2$, and temperature, due to the self-consistent parameters $e$ and $p$. At half-filling,
the spectral function for the lower Hubbard
band determines the amplitudes in $|\rm MI\rangle$
through $\sigma^{1}=(u_k+ v_k)^2/2$.  In the strong-coupling limit, $U \gg t$, the only dependence that remains is on filling. In this limit $\sigma_{\xi\xi}^{(1)} = I_1$, $\sigma_{\eta\eta}^{(2)} = I_2$ and all the other functions vanish. All of our calculations of the Hall coefficient were performed
in this limit.  

Consider now the two weak-coupling limits in which
Fermi liquid theory should hold: Case a) $n\rightarrow 0$ and Case b)
$U\rightarrow 0$. In the first case, $g\rightarrow -U+2m/n$, $Q\rightarrow |g|/2$, the correlations in $p$ become independent and hence $p\propto n^2$, $e\propto n$ implying
that $m(\vec k)\propto n$.  Consequently, the dispersion for the lower Hubbard band
reduces exactly to that of the non-interacting limit,
 $\epsilon_1(\vec k)=-\mu-4t\alpha(\vec k)=\epsilon_0(\vec k)$.  Moreover, all the spectral weight resides in this band because as $n\rightarrow 0$,
 $g/2Q=1$ and $m(\vec k) \rightarrow 0$,
implying that $\sigma^{(1)}=1$ and $\sigma^{(2)}=0$.  That the static approximation reduces to the correct non-interacting limit is not unexpected as Figures
 (1) and (2) illustrate that the Fermi surface is independent of $U$
as $n\rightarrow 0$.  In the $U\rightarrow 0$ limit, $g(\vec k)\rightarrow
(1-n)m(\vec k)/I_1I_2$ and as a consequence, $Q=|m(\vec k)|/2I_1I_2$. As a result,
the band dispersion relations are $\epsilon_{1,2}=\epsilon_0(\vec k)-(m(\vec k)\mp|m(\vec k)|)/2I_1I_2$ with spectral weights
$\sigma^{(1,2)}=1/2\pm m(\vec k)/(2|m(\vec k)|)$ which are either unity or zero.  Consequently,
although two bands still exist, only the free particle dispersion carries unit spectral weight because the $|m(\vec k)|$ terms enter with opposite signs.  Hence, the static approximation correctly reproduces
the non-interacting limit when $U\rightarrow 0$.  As a consequence,
the violation of Luttinger's theorem seen here is not an artifact of the 
approximation scheme but stems fundamentally from the splitting of the spectral
weight over two bands although no symmetries are broken, the hallmark of Mottness.

\end{document}